\pdfoutput=1
\documentclass[11pt]{article}

\usepackage{jheppub}
\usepackage{amsmath}
\usepackage{verbatim}
\usepackage{amssymb}
\usepackage{inputenc}
\usepackage{textcomp}
\usepackage{appendix}
\usepackage{mathtools}
\usepackage{multirow}
\usepackage{mathrsfs}

\newcommand{\e}{\epsilon}
\newcommand{\be}[1]{\begin{equation}\label{#1} }
\newcommand{\ee}{\end{equation}}
\newcommand{\bea}[1]{\begin{eqnarray}\label{#1} }
\newcommand{\eea}{\end{eqnarray}}
\newcommand{\bes}{\begin{subequations}}
\newcommand{\ees}{\end{subequations}}

\newcommand{\p}{\partial}

\newcommand{\refb}[1]{(\ref{#1})}

\newcommand{\Q}{{\mathcal{Q}}}

\renewcommand{\l}{\mathcal{L}}
\newcommand{\bs}{{\bar{\sigma}}}

\newcommand{\bp}{\bar{\partial}}

\newcommand{\ga}{\gamma}
\newcommand{\de}{\delta}
\newcommand{\bde}{\bar{\delta}}

\newcommand{\non}{\nonumber}

\newcommand{\vr}{\varrho}
\newcommand{\bvr}{\bar{\varrho}}
\newcommand{\bQ}{\bar Q}

\newcommand{\bS}{\bar S}
\newcommand{\da}{\dot \alpha}
\newcommand{\db}{\dot \beta}
\newcommand{\dg}{\dot \gamma}
\newcommand{\dr}{\dot{\rho}}

\newcommand{\Th}{\Theta}
\newcommand{\bTh}{\bar\Theta}
\renewcommand{\L}{\Lambda}
\newcommand{\bL}{\bar\Lambda}
\newcommand{\tG}{\tilde G}
\newcommand{\im}{\implies}

\newcommand{\D}{\Delta}

\renewcommand{\(}{\left(}
\renewcommand{\)}{\right)}

\renewcommand{\a}{\alpha}

\renewcommand{\b}{\beta}

\newcommand{\s}{\sigma}

\preprint{TUW--22--01}

\title{Carrollian superconformal theories and super BMS}

\author[a]{Arjun Bagchi,} \author[b]{Daniel Grumiller,} \author[a,c]{and Poulami Nandi.} \author{\\}

\affiliation[a]{Indian Institute of Technology Kanpur, Kalyanpur, Kanpur 208016. India. \\}
 
\affiliation[b]{Institute for Theoretical Physics, TU Wien, Wiedner Hauptstrasse 8--10, 1040 Vienna, Austria.\\}

\affiliation[c]{Indian Institute of Technology Gandhinagar, Gujarat 382355, India.\\}

\emailAdd{abagchi@iitk.ac.in, grumil@hep.itp.tuwien.ac.at, poulami.n@iitgn.ac.in}

\abstract{We investigate the supersymmetric versions of Bondi--Metzner--Sachs or, equivalently, conformal Carroll symmetry in boundary dimensions $d>3$, with applications of flat space holography in mind. We identify the contraction of the relativistic symmetry relevant for our purposes and construct a finite-dimensional Carrollian superconformal algebra (CSA) before proposing an infinite-dimensional lift. We provide the superspace formulation for $\mathcal{N}=1$ CSA and work towards an understanding of the representation theory of the algebra. We conclude with some aspects of $\mathcal{N}>1$ CSA.} 

\keywords{Carrollian superconformal field theories, super BMS, Carrollian superconformal algebra, flat space holography, AdS/CFT}

\begin{document}

\maketitle

\section{Introduction}
The Holographic Principle \cite{tHooft:1993dmi,Susskind:1994vu} is, at present, our most promising tool for understanding quantum gravity. Its best-studied realization is the Anti-de~Sitter/Conformal Field Theory (AdS/CFT) correspondence \cite{Maldacena:1997re,Gubser:1998bc,Witten:1998qj,Aharony:1999ti}. 
The idea of holography, {\it viz.} understanding a higher-dimensional gravitational theory in terms of a theory without gravity that lives on its boundary, was based on the fact that the entropy of a black hole is proportional to its area and not its volume. This area law is not tied to AdS spacetimes. Therefore, the overwhelming expectation is that holography is true in general and applies also to spacetimes that are not asymptotically AdS. In particular, an outstanding question is how to formulate a holographic duality for asymptotically flat spacetimes, see \cite{Polchinski:1999ry,Susskind:1998vk,Giddings:1999jq} for some early attempts.

\medskip

\noindent Given the spectacular success of AdS/CFT, a particularly compelling way to attempt building holography in flat space is to take the AdS radius to infinity and the corresponding limit on the CFT side. It has been shown that the infinite radius limit in the bulk manifests itself as a limit where the speed of light in the dual field theory goes to zero \cite{Bagchi:2012cy}. This ultra-relativistic (UR) limit leads to a class of field theories called Carrollian field theories, where the Carroll group replaces the relativistic Poincar{\'e} group. The conformal versions of these are Carrollian CFTs. The putative dual theories of flat space are therefore Carrollian CFTs in one lower dimension \cite{Bagchi:2010eg,Bagchi:2012cy, Bagchi:2016bcd,Bagchi:2019xfx,Bagchi:2019clu,Banerjee:2020qjj}. These field theories live naturally on the null boundary of asymptotically flat spacetime as well as on black hole horizons, see e.g.~\cite{Donnay:2019jiz}.

\medskip

\noindent
Asymptotic symmetries are generated by boundary condition preserving transformations that do not fall off sufficiently fast near the boundary, see e.g.~\cite{Compere:2018aar,Harlow:2019yfa} and refs.~therein. In many applications, the associated asymptotic symmetry group (ASG) provides an infinite enhancement of the bulk isometries. The most famous examples are asymptotically flat spacetimes in 4d, studied originally by Bondi, van~der~Burgh, Metzner and Sachs (BMS) \cite{Bondi:1962px,Sachs:1962zza}, who found an enhancement of Poincar\'e symmetries by an infinite set of angle-dependent translations, and asymptotically AdS spacetimes in 3d, where the analysis of Brown and Henneaux \cite{Brown:1986nw} uncovered two copies of the Virasoro algebra generating the asymptotic symmetries. The latter example is seen by many as a precursor to the AdS/CFT correspondence. Therefore, the BMS analysis could be a precursor to a flat space holographic correspondence. This vision is behind many current approaches to flat space holography, including ours.

\medskip

\noindent 
One version of the BMS group is the semi-direct product of the group of all (local) conformal transformations of the sphere at infinity (also known as super-rotations) with super-translations, the angle-dependent translations at null infinity.\footnote{
The original BMS analysis did not allow for super-rotations, which were introduced in \cite{Barnich:2009se}. Super-rotations can be further generalized to arbitrary diffeomorphisms of the sphere, see \cite{Campiglia:2014yka}. The physical relevance of superboosts was uncovered in \cite{Compere:2018ylh}. Moreover, super-translations can be generalized to have arbitrary spin, see the discussion in \cite{Grumiller:2019fmp,Campoleoni:2020ejn}.
}  For bulk dimension $D>4$, there are choices of boundary conditions for which these BMS groups are again infinite-dimensional \cite{Kapec:2015vwa,Fuentealba:2021yvo} (as opposed to those which keep just the Poincar{\'e} group \cite{Hollands:2003ie,Tanabe:2009va,Tanabe:2011es}). These choices seem to be the ones that are most interesting for physical purposes, keeping in mind results like Weinberg's soft graviton theorem \cite{Weinberg:1965nx} that holds in all dimensions and recent findings that link these theorems to asymptotic symmetries \cite{He:2014laa}.

\medskip

\noindent
A recipe for 
holography in arbitrary spacetimes is to compute the ASG for the bulk gravitational theory and posit that the ASG generates the global symmetries of the dual boundary field theory. Following this line of argument, the BMS group should govern the boundary dynamics of field theories putatively dual to asymptotically flat spacetimes \cite{Bagchi:2010eg, Bagchi:2012cy}. This program was implemented successfully in 3d, including a concrete proposal for a field theory dual \cite{Bagchi:2012yk}, Cardy-type of state counting from BMS-symmetries \cite{Bagchi:2012xr,Barnich:2012xq}, the holographic calculation of correlation functions \cite{Bagchi:2015wna}, entanglement entropy \cite{Bagchi:2014iea,Jiang:2017ecm,Hijano:2017eii}, and more; a selected list of papers is \cite{Barnich:2012rz,Bagchi:2013qva,Bagchi:2013lma,Barnich:2014kra,Krishnan:2013wta,Hartong:2015usd,Campoleoni:2016vsh,Bagchi:2016geg,Hijano:2017eii,Hijano:2018nhq}. In this paper, we focus on higher dimensions. Some relevant literature in this context is \cite{Bagchi:2016bcd,Bagchi:2019xfx,Bagchi:2019clu,Banerjee:2020qjj, Campoleoni:2021blr, Chen:2021xkw}. Duals of 4d flat space in terms of a 2d CFT living on the celestial sphere has gathered a lot of recent interest. We will not be exploring this rather intriguing direction in our work here. We point the interested reader to the excellent reviews \cite{Strominger:2017zoo,Pasterski:2021rjz}. 

\medskip

\noindent The limiting construction alluded to above is consistent between gravity and field theory sides. Namely, the Carrollian conformal algebra (CCA) is isomorphic to the BMS algebra in one higher dimension \cite{Duval:2014uva}.
\be{ccabms}
\mathfrak{Cconf}_d = \mathfrak{bms}_{d+1}
\ee
Thus, the limiting construction and the intrinsic construction are consistent with each other, at least at the level of the symmetries. 

\medskip

\noindent
A final point of clarification in this limiting vs.~intrinsic analyses is the infinite enhancement of symmetries. As we just mentioned, the BMS algebra is infinite-dimensional for bulk dimensions $D=3,4$ and for certain boundary conditions in $D>4$. In AdS spacetimes, on the other hand, until recently there were infinite enhancements only when $D=3$.\footnote{
In \cite{Compere:2019bua} new AdS$_4$ boundary conditions were introduced that lead to infinite-dimensional asymptotic symmetries and allow to take a flat space limit to BMS. The implications of this approach for AdS/CFT are not completely clear yet. Thus, we follow a different route.} The natural question then is how one can see these infinite enhancements for boundary dimensions $d>3$ from the point of view of the limit. It turns out that the finite algebra ($iso(d,1)$) that one gets by performing the {\.I}n\"on\"u--Wigner contraction of the $so(d,2)$ algebra of the isometries of AdS$_{d+1}$ is the Poincar{\'e} algebra in $(d+1)$ dimension, which can be lifted to an infinite-dimensional algebra. This infinite lift matches with the infinite enhancement of the BMS group.

\medskip

\noindent
Our principal long-term goal is to establish flat space holography, specifically to construct quantum field theories that are dual to (quantum) gravity in asymptotically flat spacetimes. More modestly, it would already be progress to understand one example in some detail. The best-known realization of the Holographic Principle is Maldacena's correspondence between Type IIB superstring theory on AdS$_5 \times$ S$^5$ and $\mathcal{N}=4$ super Yang--Mills theory in $d=4$ \cite{Maldacena:1997re}. It would thus be useful to understand if a version of this correspondence exists for asymptotically flat spacetimes. The bulk side is well-understood: neglecting the S$^5$ for now, in the weak coupling regime, the gravitational theory is Type IIB supergravity in $D=5$ flat spacetimes. We would like to understand the analogue of this flat space limit on the dual field theory side. In view of what we have introduced above, the putative dual would be a Carrollian superconformal field theory.\footnote{While our focus is on null infinity, it is also possible to consider
spatial infinity, where non-linear asymptotic symmetries emerge in higher dimensions \cite{Fuentealba:2021yvo}. Currently, there is no proposal for a dual field theory at spatial infinity, nor is there a physical interpretation of what such a field theory would represent.} Our objective in this paper is to initiate a study of these field theories by formulating the algebraic structure behind them. In short, we construct Carrollian superconformal symmetry and concentrate on boundary dimensions greater than three.\footnote{There are known issues regarding the definition of the null structure and radiation in asymptotically flat spacetimes in odd dimensions higher than four \cite{Valiente-Kroon:2002xys,Hollands:2004ac}. Our algebraic constructions in this paper is not sensitive to these issues.}

\medskip

\noindent
Our paper is organized as follows. In Sec.~\ref{Scaling for the Bosonic generators of Finite CCA}, we revisit earlier work on the  CCA and its representation theory. In Sec.~\ref{sec3}, we focus the $\mathcal{N}=1$  Carrollian Superconformal Algebra (CSA). We discuss types of possible scaling that generate the algebra and then construct a superspace formulation for it. Later in the section, we extend the finite $\mathcal{N}=1$ CSA to admit the infinite `fermionic super-translations' in $d=4$ and $R$-symmetry. We generalise the infinite CSA to $\mathcal{N}=4$ in Sec.~\ref{n=4}. The proposed infinite-dimensional lift is consistent with results in the existing literature for $d=2, 3$. In Sec.~\ref{sec8}, we work out aspects of the representation theory of the infinite CSA from the intrinsic point of view, by considering the actions of the generators on Carrollian fields. The obtained intrinsic multiplet structure matches that from the limiting picture. We finish in Sec.~\ref{Conclusions} with a summary and discussions of other current and future directions of research. 

\medskip

\noindent
There are four appendices consisting of details omitted in the main text. 
Appendix \ref{relal} displays the most useful form of the relativistic $SU(2,2|N)$ algebra for our purposes. Appendix \ref{appenb} provides a consistency check for our proposed infinite-dimensional extension of the Carrollian superconformal symmetries by reproducing various results on super BMS algebras that have recently appeared in the literature. Appendix \ref{identities section} lists identities that have been used throughout the paper. Appendix \ref{appen3} shows all the details of the representation theory of infinite CSA. 

\section{Bosonic story --- Carrollian CFTs}\label{Scaling for the Bosonic generators of Finite CCA}
We begin our discussions by revisiting the bosonic construction of Carrollian CFTs. We shall quickly remind the reader of the UR limit of a CFT in generic dimensions and the corresponding finite-dimensional CCA emerging from it. We then give this an infinite lift. Finally, mention some rudimentary facts about Carrollian CFT representation theory. 

\subsection{Algebra} 
The UR or Carrollian limit of a relativistic CFT is reached by performing  an In\"{o}n\"{u}--Wigner
contraction on the relativistic conformal generators. The corresponding contraction of the spacetime coordinates for a $d$-dimensional CFT is described as
\be{stscale}
x_i \to x_i\qquad\qquad t \to \e\, t\qquad\qquad \e \to 0\,.
\ee
Here, $i$ runs over the spatial coordinates $i=1,\hdots,d-1$. The above contraction can also be interpreted as taking the limit of vanishing speed of light, $c\to 0$. The UR generators are obtained by performing the space-time contraction on the parent relativistic generators. For example, we obtain UR boost generator by scaling  relativistic boost and regularising it as
\be{convensca}
B_i^{\textrm{\tiny rel}}=  -x_i \p_t-t\p_i \xrightarrow[]{\text{rescale}\, t} -\frac{1}{\e}x_i \p_t-t\p_i 
 \xrightarrow[]{\text{rescale}\, B_i}  B_i= 
\displaystyle \lim_{\e \rightarrow 0}\e B_i^{\textrm{\tiny rel}} 
 \xrightarrow[\text{limit}]{\text{Carroll}} B_i=- x_i \p_t\,.
\ee
The other UR generators are obtained similarly from their parent relativistic conformal generators. They are given by
\begin{subequations}
\label{genearl}
\begin{align}
H &= \p_t &  B_i&=-x_i \p_t &  K_i &= -2 x_j (t\p_t+x_i\p_i)+x_j x_j \p_i & K &=x_i x_i \p_t \\
 D&=-(t\p_t+x_i \p_i) & P_i&=\p_i & J_{ij}&=-(x_i\p_j-x_j\p_i)\,. &&
\end{align} 
\end{subequations}
These generate the finite Conformal Carrollian Algebra (f-CCA),\footnote{We are particularly interested in level-$N$ CCA (CCA$_N$)  when $N=2$. The UR limit of the conformal generators gives rise to generators of CCA$_N$ for $N=2$. For details see \cite{Duval:2014uva,Duval:2014lpa,Bagchi:2019clu}.} which is $iso(d,1)$ for a $d$-dimensional field theory \cite{Bagchi:2016bcd,Bagchi:2019xfx}.
\begin{subequations}
\label{algebra}
\begin{align}
[J_{ij}, B_k ]&=\delta_{k[i}B_{j]} & [J_{ij}, P_k ]&=\delta_{k[i}P_{j]} & [J_{ij}, K_k ]&=\delta_{k[i}K_{j]} & [B_i,P_j]&=\delta_{ij}H\\
[B_i,K_j]&=\delta_{ij}K & [D,K]&=-K & [K,P_i]&=2B_i & [K_i,P_j]&=-2D\delta_{ij}-2J_{ij} \nonumber \\ [H,K_i]&=2B_i & [D,H]&=H & [D,P_i]&=P_i & [D,K_i]&=-K_i\,.
\end{align}
\end{subequations}
We focus on $d=4$. The sub-algebra consisting of the generators $\{J_{ij}, B_i, P_i, H\}$ forms the $c\to0$ limit of the Poincar{\'e} algebra {\it viz.} the Carrollian algebra \cite{Leblond65}. The generators $\{J_{ij},P_i,D,K_i\}$ form  $so(5)$, the conformal algebra of $3$-dimensional Euclidean space. 

\medskip

\noindent
We use an equivalent way of writing the generators of f-CCA  from their parent relativistic generators :
\be{bosgen}
\{H^{\text{rel}}, K^{\text{rel}}, B^{\text{rel}}_i\}= \frac{1}{\e}\{H, K, B_i\} \quad K^{\text{rel}}_i = K_i\quad P^{\text{rel}}_i= P_i \quad D^{\text{rel}}= D \quad J_{ij}^{\text{rel}} = J_{ij}
\ee
We rephrase this by saying
\be{1}
G_A \to \frac{1}{\e} \,G_A\qquad\qquad  G_a \to  G_a
\ee
where $G_A$ is the set of generators that scale in the limit $c\to0$, while $G_a$ is the set that remains unchanged. 
It is possible to give the finite algebra in \eqref{algebra} an infinite-dimensional lift by introducing time translation generator with arbitrary spatial dependence
\be{2}
M_f=f(x_i)\p_t\,.
\ee
Here, $M_f$ generates the infinite set of super-translations and  $f(x_i)$ is an arbitrary function of the spatial co-ordinates $x_i$, which we restrict to polynomials. We obtain the finite generators of f-CCA, i.e., $M_f = H,B_i,K$ when $f(x_i)=1,-x_i,x_k x_k$ respectively.  The super-translation generators $M_f$ along with the finite set of generators $\{B_i,J_{ij},H,P_i,D,K,K_i\}$  describe the infinite-dimensional CCA. For $d\geq 4$ it can be written as \cite{Basu:2018dub,Bagchi:2016bcd}:
\begin{subequations}
\label{infinitealgebra1}
\begin{align}
[P_i, M_f] &=M_{\p_i f} & \quad [D,M_f]  &=M_{(-x_i \p_i f+f)}\\
[K_i,M_f]&= M_{2x_i f+x_k x_k\p_i f-2x_ix_k\p_k f} &\quad [J_{ij},M_f]&= M_{-x_{[i}\p_{j]}f}\,.
\end{align}
\end{subequations}
For a more explicit version of the algebra (in terms of modes), the reader is pointed to \cite{Bagchi:2016bcd}.

\subsection{Representation theory} 
In \cite{Bagchi:2016bcd},  the representation theory of the CCA based on highest weights was described. The analysis was further extended to fields of different integer and half-integer spins in \cite{Bagchi:2019xfx}. Even though the construction was primarily intended for $d=4$, it was expected to work for higher dimensions as well. For the CCA, the states were labeled with the eigenvalues of dilatation and rotation generators. We briefly describe this construction below.

\medskip

\noindent 
We label the Carrollian CFT fields with scaling dimension $\Delta$ and spin $j$ as
\be{4}
[D,\Phi(0,0)]=\Delta \Phi(0,0)\qquad\qquad [J^2, \Phi(0,0)]=j(j+1)\Phi(0,0)\,.
\ee 
The action of Carrollian rotation, space- and time-translation on a generic field is given by
\be{5}
[J_{ij},\Phi(0,0)]=\Sigma_{ij}\Phi(0,0),\quad[H,\Phi(t,x_i)]=\p_t \Phi(t,x_i),\quad[P_i,\Phi(t,x_i)]=\p_i \Phi(t,x_i)\,.
\ee
The Carrollian conformal primaries are defined as 
\be{6}
[K_i,\Phi(0,0)]=0,\;\; [K,\Phi(0,0)]=0,~~[M_{f},\Phi(0,0)]=0~~\text{for polynomial degree} > 1\,.
\ee
The fields are not eigenstates of Carrollian boosts. Hence, using the Jacobi identity, the transformation of a generic field under Carrollian boosts can be written as
\be{7}
[B_k,\Phi(0,0)]=r\varphi_k+\, f \sigma_k\phi + f^{\prime} \sigma_k\chi\, + a A_t \delta_{ik}+b A_k+ \hdots
\ee
Here, $\varphi, \,\{\phi,\chi \},\,  \{A_t , A_k\}$ denote the primaries involving different spins $(0,\frac{1}{2},1)$. The constants $r,\{f,f^\prime\},\{a,b\}$ are    determined  from the dynamics of the corresponding theory. The above expression can be generalised for higher spins as well.

\medskip

\noindent Using the fact that the Carrollian primary $\Phi(t,x_i)$ evolves to a generic spacetime point from the origin as 
\be{8}
\Phi(t,x)=U \Phi(0,0) U^{-1}\quad \text{where} ~ U=e^{-tH-x_i P_i}
\ee
the action of the finite and infinite sets of generators of CCA on this generic Carrollian primary $\Phi(t,x_i)$ can be written as
\begin{subequations}
\label{repgen}
\begin{align}
  [J_{ij}, \Phi(t,x_i)]
&=- (x_i \p_j-x_j \p_i ) \Phi(t,x)+\Sigma_{ij}\Phi(t,x_i)\\
[B_j, \Phi(t,x_i)]
&=-x_j\p_t \Phi(t,x)+U_j\qquad\qquad\qquad\qquad U_j:=U [B_j, \Phi(0,0)]U^{-1}\\ 
[D, \Phi(t,x_i) ]
&= (-t\p_t-x_i \p_i+\D) \Phi(t,x_i)\\ 
 [K_j, \Phi(t,x_i) ]
&= - (2\Delta x_j+2x_jt\partial_t+2x_i x_j \partial_i-2x_i \Sigma_{ij}- x_i x_i \partial_j )\,\Phi(t,x)+2t \, U_j\\ 
[M_f, \Phi(t,x) ] &= f(x_i)\p_t \Phi(t,x)+\p_j f\:U_j\,.
\end{align}
\end{subequations}
In later sections, we will use these results from the bosonic representation theory as hints to build the representations for the supersymmetric version. 

\section{\texorpdfstring{$\boldsymbol{\mathcal{N}=1}$}{N=1} Carrollian superconformal symmetry}\label{sec3}
Having acquainted ourselves with the basics of the bosonic construction, we now venture into the world of Carrollian superconformal symmetries. We begin with the assertion 
\be{CB}
\mathfrak{Csconf}_d = \mathfrak{sbms}_{d+1}.
\ee
This is a natural supersymmetric generalisation of the earlier isomphism \refb{ccabms}. In this section, we construct CSAs by taking UR contractions of the parent relativistic superconformal algebra and then from the suggestive form of the contracted algebras, give these infinite-dimensional lifts. Supersymmetric BMS algebras have been studied in the literature for lower dimensions. For the known cases of $d=2$ \cite{Barnich:2014cwa,Banerjee:2016nio,Lodato:2016alv,Bagchi:2016yyf,Caroca:2018obf,Fuentealba:2017fck} (and $d=3$ \cite{Fotopoulos:2020bqj}), we shall check our proposal \refb{CB}. 

There is an important point we clarify first. The $d=2$ algebra ($\mathfrak{sbms}_3$) has two distinct sub-categories, called the homogeneous and inhomogeneous \cite{Bagchi:2015nca,Bagchi:2017cte} (or democratic/despotic \cite{Lodato:2016alv}). These correspond to asymptotic symmetries of normal supergravity and twisted supergravity respectively. For our higher-dimensional explorations in this paper, from the context of holography of asymptotically flat spacetimes, at present, we are interested in the dual to normal supergravity and hence focus on the homogeneous algebras. The contractions we perform are chosen to achieve this. We leave the inhomogeneous algebras for future work. Below we first give a brief account of these two different types of super-BMS algebras.

\subsection{Two different SUSY extensions}\label{sec:3.1}
The BMS algebras arise as the asymptotic symmetries of gravitational theories at the null boundary of asymptotically flat spacetimes. The super BMS algebras, in the same way, are expected to arise as the asymptotic symmetries of flat space supergravity. The asymptotic symmetries of the minimal supersymmetric extension to 3D Einstein gravity, viz. $\mathcal{N}=1$ supergravity, was constructed in \cite{Barnich:2014cwa}. The resulting asymptotic symmetry algebra is  
\begin{subequations}
\label{n1}
\begin{align}
[L_n,L_m]&=(n-m)\,L_{n+m}+\frac{c_L}{12}\,n(n^2-1)\delta_{n+m,0}\\
[L_n,M_m]&=(n-m)\,M_{n+m}+\frac{c_M}{12}\,n(n^2-1)\delta_{n+m,0}\\
[L_n,Q_r]&=(\frac{n}{2}-r)\,Q_{n+r}\\
\{Q_r,Q_s\}&=M_{r+s}+\frac{c_M}{6}\,\big(r^2-\tfrac{1}{4}\big)\,\delta_{r+s,0}\,.
\end{align}
\end{subequations}
Here, $L_n,M_n$'s generate the super-rotations and super-translations respectively, and the fermionic generators are given by $Q_r$, with $n,m\in\mathbb{Z}$ and either $r,s\in\mathbb{Z}$ (Ramond) or $r,s\in\mathbb{Z}+\frac12$ (Neveu--Schwarz). For Poincar{\'e} supergravity with Newton's constant $G$, we get $c_L=0, c_M=\frac{3}{G}$. The above algebra can be obtained by starting with a super-Virasoro algebra $(\l^+_n, \Q^+_r)$ and a Virasoro algebra $(\l^-_n)$
\begin{subequations}
\label{11a}
\begin{align}
[\l^\pm_n, \l^\pm_m] &= (n-m)\, \l^\pm_{n+m} +\frac{c^\pm}{12}\,n(n^2-1)\delta_{n+m,0}\\
[\l^+_n, \Q^+_r] &= \big(\frac{n}{2}-r\big)\, \Q_{n+r}\\
\{\Q^+_r, \Q^+_s\} &= 2 \l_{r+s} +\frac{c^+}{3} \,\big(r^2-\tfrac{1}{4}\big)
\end{align}
\end{subequations}
and contracting it as follows
\be{9}
L_n = \l^+_n - \l^-_{-n}\qquad\qquad M_n = \e\, (\l^+_n + \l^-_{-n})\qquad \qquad Q_r = \sqrt{\e} \,\Q^+_r\,.
\ee
When there are more supersymmetry generators, there is more room to man{\oe}uvre and one can scale the relativistic superalgebra in different ways, keeping the bosonic part of the algebra the same and thus obtaining different supersymmetric extensions of the contracted algebra \cite{Bagchi:2016yyf}. 

We now focus on the relativistic $\mathcal{N}=(1,1)$ case. In \cite{Lodato:2016alv}, it was shown that there are two distinct theories of supergravity in asymptotically flat spacetimes, with two different asymptotic symmetry algebras, that one can obtain starting out with $\mathcal{N}=(1,1)$ supergravity in AdS$_3$. These two different symmetry algebras also arise on the worldsheet of the tensionless superstring \cite{Bagchi:2016yyf,Bagchi:2017cte, Bagchi:2018wsn}. The homogeneous or democratic limit leads to an analogue of the above algebra \refb{n1}:  
\begin{subequations}
\label{noR}
\begin{align}
[L_n,L_m]&=(n-m)\,L_{n+m}+\frac{c_L}{12}\,n(n^2-1)\delta_{n+m,0}\\
[L_n,M_m]&=(n-m)\,M_{n+m}+\frac{c_M}{12}\,n(n^2-1)\delta_{n+m,0}\\
[L_n,Q^{\a}_r]&=\Big(\frac{n}{2}-r\Big)\,Q^{\a}_{n+r}\\
\{Q^{\a}_r,Q^{\b}_s\}&=\delta^{\a \b}\,\Big(M_{r+s}+\frac{c_M}{6}\,\big(r^2-\tfrac{1}{4}\big)\,\delta_{r+s,0}\Big)\,.
\end{align}
\end{subequations}
Here, we have $\a,\b=+,-$. So this differs from the previous algebra only in terms of the number of fermionic generators. The contraction, from two copies of the super-Virasoro algebra, now becomes
\be{10}
L_n = \l^+_n - \l^-_{-n}\qquad \qquad M_n = \e \,(\l^+_n + \l^-_{-n})\qquad \qquad Q^\pm_r = \sqrt{\e}\, \Q^\pm_r\,.
\ee
The supercharges scale in the same way and hence the name homogeneous. One can also start with a super-Virasoro with additional $R$-symmetry and a Virasoro and contract to get $R$-symmetry in the super BMS algebra. Some details of such algebras and ones with higher $\mathcal{N}$ are presented in the appendix. The principal point to note in the above algebras \refb{n1} and \refb{noR} is that the anticommutator between the fermionic generators always produce super-translations $M_n$. This is what will distinguish this sector from its inhomogeneous counterpart. 

We now elaborate on the inhomogeneous version of super BMS. Here, unlike the homogeneous sector, we take different combinations of the fermionic supercharges and scale them asymmetrically. The contraction that gets one from two copies of super-Virasoro to the inhomogeneous BMS algebra is 
\be{11}
L_n = \l^+_n - \l^-_{-n}\,, \quad M_n = \e (\l^+_n + \l^-_{-n})\,, \quad G_r = \Q^+_r - i \Q^-_{-r}\,, \quad 
\bar{G}_r = \e(\Q^+_r + i \Q^-_{-r})\,.
\ee
The anti-commutators of the fermionic generators $(G_r,\bar{G}_s)$ now produce the super-rotations $L_n$ as well as the super-translations $M_n$. The inhomogeneous or despotic super BMS$_3$ algebra is obtained as a flat space limit to $\mathcal{N}=(1,1)$ AdS$_3$ supergravity \cite{Lodato:2016alv}. The resulting supergravity theory is a twisted version of flat space supergravity, the asymptotic symmetries of which are given by
\begin{subequations}
\label{1a}
\begin{align}
[L_m,L_n]&=(m-n)\,L_{m+n} & [L_m,M_n]&=(m-n)\,M_{m+n}\\
[L_m,G_r]&=\Big(\frac{m}{2}-r\Big)\,G_{m+r} &[L_m,\bar{G}_r]&=[M_m,G_r]=\(\frac{m}{2}-r\)\,\bar{G}_{m+r} \\
\{G_r,G_s\}&=2L_{r+s}+\frac{c_L}{3}\,\big(r^2-\tfrac{1}{4}\big)\,\delta_{r+s,0} & \{G_r,\bar{G}_s\}&=2M_{r+s}+\frac{c_M}{3}\,\big(r^2-\tfrac{1}{4}\big)\,\delta_{r+s,0}\,.
\end{align}
\end{subequations}
For the purposes of this paper, we are interested in constructing higher-dimensional version of the super BMS or equivalently CSAs. We focus on the symmetries of the dual field theory of Poincar{\'e} supergravity and hence concentrate entirely on the homogeneous versions of the above-mentioned algebras.

\subsection[Constructing  the \texorpdfstring{$\mathcal{N}=1$}{N=1} algebra]{Constructing  the \texorpdfstring{$\boldsymbol{\mathcal{N}=1}$}{N=1} algebra}
We aim to construct CSAs for arbitrary dimensions. For the moment, we focus on $d=4$. The strategy that we follow is the following: 
\begin{itemize}
\item Take an appropriate UR contraction of the relativistic superalgebra $SU(2,2|\mathcal{N})$, so that we reach the homogeneous version of the contracted algebra. This generates a finite algebra with the same number of generators as $SU(2,2|\mathcal{N})$. 
\item Write the super-algebra in a suggestive form and give it an infinite-dimensional lift. For this, we will construct a Carrollian superspace and will write the generators in superspace coordinates.  
\item Check the consistency with similar constructions for the lower-dimensional cases to verify the isomorphism with the super BMS algebras. 
\end{itemize}
We begin our analysis with the simplest case of $\mathcal{N}=1$ supersymmetry. For $\mathcal{N}=1$, the number of supercharges in the parent relativistic theory is 8 (4 supercharges $\{Q,\bar{Q}\}$, and 4 superconformal charges $\{S,\bar{S}\}$):  
$
(Q_\a, \bar Q_{\dot \a},S_\a,  \bar S_{\dot \a}),~\text{where} \:\a,\dot{\a}=1,2.
$
We also have the $R$-symmetry group, which is $U(1)$ for $\mathcal{N}=1$. There can be numerous ways to perform the UR contraction on the relativistic superconformal algebra, resulting in different possible algebras. As explained above, taking a cue from the two-dimensional algebras, we focus on the homogeneous algebras, where the fermionic generators close to form super-translations. 

\subsubsection*{Possible scalings}
We start by considering two ways to contract the fermionic generators of $\mathcal{N}=1$ relativistic superconformal algebra in $d=4$, namely the symmetric and asymmetric scaling. Both will lead to the same algebra.  

\subsubsection*{Symmetric scaling}\label{ss}
First, we describe the symmetric scaling of the 8 supercharges. We assume
\be{assum}
 \bar{Q}_{\dot \alpha}= (Q_\a)^\dagger\qquad\qquad \bar{S}_{\dot \alpha}= (S_\a)^\dagger\,.
\ee
The symmetric scaling is given by  
\be{symscale}
Q_\a \to \frac{1}{ \sqrt \e} Q_\a\qquad \bar Q_{\dot \a} \to \frac{1}{ \sqrt\e} \bar Q_{\dot\a}\qquad S_\a \to \frac{1}{ \sqrt\e} S_\a\qquad \bar S_{\dot \a} \to \frac{1}{ \sqrt\e} \bar S_{\dot\a}\qquad R \to  R\,.
\ee
The above scaling for the fermionic generators along with the bosonic generators in \eqref{bosgen} results in the following algebra:

\smallskip
\noindent
\textit{Fermionic sector:}
\begin{subequations}
\label{fer1}
\begin{align}
 \{Q_\a, \bar Q_{\dot \a} \}&=2  \sigma^0_{\a \dot \a}H & \{Q_\a,  Q_{ \b} \}&=\{\bar{Q}_{\dot \a},  \bar Q_{\dot \b} \}=0\\
 \{S_\a, \bar S_{\dot \a} \}&=2  \sigma^0_{\a \dot \a}K & \{S_\a,  S_{ \b} \}&=\{\bar{S}_{\dot \a},  \bar S_{\dot \b} \}=0\\
 \{Q_\a, S_{\b} \}&= 2 (\sigma^{0i})_\a^{~ \gamma} \e_{\gamma \b}B_i &  \{Q^A_\a, \bar S^B_{\dot \b} \}&=0 \\
 \{\bar Q_{\dot\a}, \bar{S}_{\dot\b} \}&= -2\e_{\dot \a \dot \gamma} (\bar \sigma^{0i})^{\dot \gamma}_ {~\dot \b} B_i &  \{\bar Q^A_{\dot\a}, S^B_{ \b} \}&=0
\end{align}
\end{subequations}
\smallskip
\noindent
\textit{Mixed sector:}
\begin{subequations}
\label{mix11}
\begin{align}
  [ Q_\a,P_i]&=[Q_\a, H]=0 & [ S_\a,P_i]&=i (\sigma^i)_{\a \dot \a}\bar Q^{\dot \a}\\
  [ \bar Q_{\dot\a},P_i]&=[\bar Q_\a, H]=0 & [ \bar S_{\dot\a},P_i]&=i \e_{\dot \a \dot \b}(\bar\sigma^i)^{\dot \b \a} Q_{ \a}\\
 [Q_\a, B_i]&=[\bar Q_{\dot\a}, B_i]=0 & [S_\a, H]&=[\bar S_{\dot \a}, H]=0\\
 [ Q_\a,K]&=[\bar Q_{\dot\a}, K]=0 & [S_\a, B_i]&=[\bar S_{\dot\a}, B_i]=0\\
 [ Q_\a,J_{ij}]&=-\frac{i}{2}(\sigma_{ij})^{\:\;\:\b}_\a Q_{\b} &  [ S_\a,K]&=[\bar S_{\dot\a}, K]=0\\
 [ \bar Q_{\dot\a},J_{ij}]&=-\frac{i}{2}\e_{\dot \a \dot \b}(\bar \sigma_{ij})^{\dot \b}_{\:\;\:\dot \gamma} \bar Q^{ \dot \gamma} &  [ S_\a,K_i]&=[\bar S_{\dot\a}, K_i]=0\\
 [ Q_\a,D]&=-\frac{1}{2}Q_\a & [ S_\a,J_{ij}]&=-\frac{i}{2}(\sigma_{ij})^{\:\;\:\b}_\a S_{\b}\\
 [ \bar Q_{\dot\a},D]&=-\frac{1}{2}\bar Q_{\dot\a} &  [ \bar S_{\dot\a},J_{ij}]&=-\frac{i}{2}\e_{\dot \a \dot \b}(\bar \sigma_{ij})^{\dot \b}_{\:\;\:\dot \gamma} \bar S^{ \dot \gamma}\\
 [ Q_\a,K_i]&=i (\sigma^i)_{\a \dot \a}\bar S^{\dot \a} & [ S_\a,D]&=\frac{1}{2}S_\a\\
 [ \bar Q_{\dot\a},K_i]&=i \e_{\dot \a \dot \b}(\bar\sigma^i)^{\dot \b \a} S_{ \a} &  [ \bar S_{\dot\a},D]&=\frac{1}{2}\bar S_{\dot\a}
 \end{align}
\end{subequations}
\smallskip
\noindent
\textit{$R$-symmetry sector:}
\be{ral1}
 [R,Q_\a]=-i\frac{3}{4}Q_\a\qquad [R,\bar Q_{\dot \a}]=i\frac{3}{4}\bar Q_{\dot \a}\qquad [R,S_\a]=i\frac{3}{4}S_\a\qquad [R,\bar S_\a]=-i\frac{3}{4}\bar S_{\dot\a}\,.
 \ee

\noindent The above algebra \eqref{fer1}--\eqref{ral1} is the finite part of $\mathcal{N}=1$ CSA in $d=4$. The generators $\{B_i, K, H, Q_\a, \bar Q_{\dot \a}, S_\a,\bar S_{\dot \a}\}$  are analogous to the translational part of  finite CSA. In other words, the anti commutators  \eqref{fer1} between the fermionic generators give rise to the bosonic generators $(H,B_i,K)$, which constitute the finite super-translations of the bosonic CCA.  

\smallskip

\noindent
With the $R$-symmetry scaling as in \refb{symscale}, $R$ does not appear on the RHS of the fermionic sector of the algebra and also is not a central term. However, if we choose the scaling such that  $R \to \frac{1}{\e}R$, then $R$ becomes a central term: it appears on the RHS of the fermionic sector of the algebra. The corresponding brackets are
\begin{subequations}
 \label{rdiffscal}
 \begin{align}
 \{Q_\a, \bar Q_{\dot \a} \}&=2  \sigma^0_{\a \dot \a}H & \{Q_\a,  Q_{ \b} \}&=\{\bar{Q}_{\dot \a},  \bar Q_{\dot \b} \}=0\\
 \{S_\a, \bar S_{\dot \a} \}&=2  \sigma^0_{\a \dot \a}K & \{S_\a,  S_{ \b} \}&=\{\bar{S}_{\dot \a},  \bar S_{\dot \b} \}=0\\
 \{Q_\a, S_{\b} \}&= 2 (\sigma^{0i})_\a^{~ \gamma} \e_{\gamma \b}B_i -4i \e_{\a \b}R &  \{Q^A_\a, \bar S^B_{\dot \b} \}&=0\\
 \{\bar Q_{\dot\a}, \bar{S}_{\dot\b} \}&= -2\e_{\dot \a \dot \gamma} (\bar \sigma^{0i})^{\dot \gamma}_ {~\dot \b} B_i +4i \e_{\dot\a \dot\b}R &  \{\bar Q^A_{\dot\a}, S^B_{ \b} \}&=0\\
 [Q_\a,R]&=[\bar Q_{\dot\a},R]=0 & [S_\a,R]&=[\bar S_{\dot\a},R]=0\,.
 \end{align}
\end{subequations}
 There is no change in the mixed sector.
 The scaling ($R \to \e^n R, n>0$) will result in the absence of the $R$ term on the RHS of the algebra. In this scaling, the LHS of $[Q_\a,R]$ scales as $\e^{n-1}$ and thus vanishes in the limit $\e \to 0$. 

\subsubsection*{Asymmetric scaling} \label{asymmetricscaling}
Now, we explore another type of scaling where we scale the fermionic generators aysmmetrically. Contrary to expectations, we end up with the same algebra as before. The important point here is that the hermiticity conditions change as compared to \eqref{assum}. The asymmetric scaling is given by either of the following two possibilities.
\begin{subequations}
\label{assym3}
\begin{align}
&\textrm{either:}& Q_\a &\to Q_\a & \bQ_{\da}&\to \frac{1}{\e}\bQ_{\da} & S_\a &\to \frac{1}{\e} S_\a & \bS_{\da}&\to \bS_{\da} & R&\to R\\
&\textrm{or:}&Q_\a &\to  \frac{1}{\e}Q_\a & \bQ_{\da}&\to\bQ_{\da} & S_\a &\to  S_\a & \bS_{\da}&\to\frac{1}{\e}\bS_{\da} & R&\to R\,.
\end{align}
\end{subequations}
The above scaling leaves the algebra  the same as \eqref{fer1} -- \eqref{ral1}.
However, for this case,  the hermiticity conditions
\be{12}
 \bar{Q}_{\dot \alpha}= (S_\a)^\dagger\qquad\qquad \bar{S}_{\dot \alpha}= (Q_\a)^\dagger
 \ee 
 exchange positive and negative scaling weights, analogous to usual Virasoro generators $L_n^\dagger = L_{-n}$. We elucidate the connection between the symmetric and asymmetric scaling more prominently when addressing the superspace formalism below.

\subsection{Carrollian superspace}\label{sec4}
We have explored the various types of possible scaling for the fermionic generators for $\mathcal{N}=1$ finite CSA by contracting their parent relativistic generators. We found that the algebra remains the same for both symmetric and asymmetric scalings. We now build the Carroll equivalent of the superspace formalism and connect the different scalings. In the superspace coordinates $\theta,\bar{\theta}$ the relativistic supercharges are given as
\begin{subequations}
\begin{align}
    \label{q1}
    Q_\a&=\p_\a-\s^\mu_{\a\db}\bTh^{\db}\p_\mu\\
\label{q2}\bQ_{\da} &= -\bp_{\da}+ \Th^\b\s^\mu_{\b\da}\p_\mu
\end{align}
\end{subequations}
where, $\p_\a=\frac{\p}{\p \Th^\a},\bp_{\da}=\frac{\p}{\p \Th^{\da}} $.
Let us now take the UR scaling on the superspace coordinates,
\be{thetaa}
\Th^\a \to \e^a \Th^\a\qquad\qquad \bTh^{\da} \to \e^b \bTh^{\da} \,.
\ee 
Here, $a,b$ are some arbitrary parameters which will be determined later. Now, the LHS of \eqref{q1} gives \be{13}\e^{-a} \p_\a-\e^{b-1}\s^0_{\a \db}\bTh^{\db}\p_t-\e^b \s^i_{\a \db}\bTh^{\db} \p_i.\ee 
Requiring $\{Q,\bar{Q}\}\sim H$ obtains  
\be{avalue}
b=1-a\,.
\ee
From the above relation, we can consider three simple choices
out of the infinitely many possible choices for $a$. There are two natural choices (assuming $a=b$, so that both sectors scale in the same
way or assuming $a=0$, so
that only one sector scales) and a third one related to one of them by
exchanging unbarred and barred sectors (assuming $a=1$ which is the same after exchange in $a=0$), as
evident from \eqref{thetaa}.

\subsection*{Case 1: Symmetric scaling}
Here we have $a=\frac{1}{2},\: \Th^\a \to \sqrt\e \Th^\a, \: \: \bTh^{\da} \to \sqrt\e \bTh^{\da}$. 
We first choose $a=\frac{1}{2}$, yielding 
\be{14}
Q_\a \to \frac{1}{\sqrt \e} Q_\a\qquad\qquad\bQ_{\da} \to \frac{1}{\sqrt \e}\bQ_{\da}\,. \ee
This reproduces the symmetric scaling (see section \ref{ss}).  Next we write the bosonic and other fermionic  generators in terms of the superspace coordinates. 
\paragraph{Fermionic generators:} The $Q$ supercharges in the contracted superspace are given by
\be{15}
Q_\a=\p_\a-\s^0_{\a\db}\bTh^{\db}\p_t\qquad\qquad \bQ_{\da} = - \bp_{\da}+ \Th^\b \s^0_{\b\da}\p_t\,.
\ee
The relativistic $S$ supercharges are given by 
\begin{subequations}
\label{16}
\begin{align}
S_\a&=-i \e^{\dot\b \dot\gamma} (\s_\mu)_{\a\dot\gamma }(x^\mu_{-})\bp_{\db}+i \e^{\db \dot\gamma} (\s_\mu)_{\a \dot \gamma}(x^\mu_{+})\Th^\b \s^\nu_{\b \dot \b}\p_\nu-2i (\Th \Th)\p_\a\\
\bS_{\da}&=-i \e^{\b \gamma} (\s_\mu)_{ \gamma\dot\a}(x^\mu_{+})\p_{\b}+i \e^{\b \gamma} (\s_\mu)_{\gamma \dot \a}(x^\mu_{-})\bar\Th^{\db} \s^\nu_{\b \dot \b}\p_{\nu} +2i (\bTh \bTh)\bp_{\da}\,.
\end{align}
\end{subequations}
Taking the UR limit on the first equation,
\be{17}
S_\a=\frac{1}{\sqrt \e}[-i \e^{\db \dot\gamma} (\s_i)_{\a \dot \gamma}(x_i)\bp_{\db}+i \e^{\db \dot\gamma} (\s_i)_{\a \dot \gamma}(x_i)\Th^\b \s^0_{\b \dot \b}\p_t]+\mathcal O(\sqrt \e)
\ee
is consistent with our scaling in \refb{ss}, which gives
\be{18}S_\a \to \frac{1}{\sqrt \e}S_\a\qquad\qquad\bar S_{\dot \a} \to \frac{1}{ \sqrt\e} \bar S_{\dot\a} \ee 
 The same argument holds for the second equation as well. Hence, we can write the contracted $S$ supercharges as
\begin{subequations}
\label{19}
\begin{align}
S_\a&=-i \e^{\db \dot\gamma} (\s_i)_{\a \dot \gamma}(x_i)\bp_{\db}+i \e^{\db \dot\gamma} (\s_i)_{\a \dot \gamma}(x_i)\Th^\b \s^0_{\b \dot \b}\p_t\\
\bS_{\da}&=-i \e^{\b \gamma} (\s_i)_{ \gamma\dot \a}(x_i)\p_{\b}+i \e^{\b \gamma} (\s_i)_{ \gamma\dot \a}(x_i)\bar\Th^{\dot\b} \s^0_{\b \dot \b}\p_t\,.
\end{align}
\end{subequations}

\paragraph{Bosonic generators:} The space and time translation generators do not have any fermionic pieces. 
Hence, the contracted generators in the superspace are
\be{20}
P_i=\p_i\qquad\qquad H=\p_t\,.
\ee
Taking the UR limit for the relativistic dilatation generator,
\be{21}
D=-(x_k \p_k+t\p_t+\frac{1}{2}(\Th^\a \p_\a+\bTh^{\da}\bp_{\dot \a}))\xrightarrow{\text{UR limit}}-(x_k \p_k+t\p_t+\frac{1}{2}(\Th^\a \p_\a+\bTh^{\da}\bp_{\dot \a}))
\ee
yields the superspace Carrollian dilation generator, 
\be{22}
D=-(x_k \p_k+t\p_t+\frac{1}{2}(\Th^\a \p_\a+\bTh^{\da}\bp_{\dot \a}))\,.
\ee
The relativistic SCT generators in superspace are 
\be{23}
K_\mu=\left( x^\nu x_\nu+2 (\Th \Th)(\bTh \bTh)\right)\p_\mu-2\left(x_\mu x^\nu + (\Th \s_\mu \bTh)(\Th \s^\nu\bTh)\right)\p_\nu-\e^{\b \gamma}(\s^\nu \bar \sigma_\mu)_\gamma^{\;\alpha} \Th_\a x_\nu \p_\b\,.
\ee
First, consider the temporal part of SCT which we call $K$. Taking the UR limit, the contracted $K$ in superspace is given by
\be{24}
K= x_i x_i \p_t\,.
\ee 
The spatial part of SCT $K_i$ contracts to
\be{25}
K_i=-2x_i (t\p_t+x_k \p_k) +x_k x_k \p_i-\e^{\b \gamma}(\s^j \bar \sigma_i)_\gamma^{\;\alpha} \Th_\a x_j \p_\b\,.
\ee
The relativistic Lorentz generators are
\be{boostrot}
J_{\mu\nu}=-x_\mu \p_\nu+x_\nu \p_\mu-\frac{i}{2}\(\s^{\a \b}_{\mu \nu}\Th_\a \p_\b-\bar\s^{\da\db}_{\mu \nu}\bTh_{\da} \bar{\partial}_{\db}\)\,.
\ee
In the UR limit, these give rise to Carrollian boosts and rotations in superspace
\be{26}
B_i=-x_i\p_t\qquad\qquad J_{ij}=-x_i \p_j+x_j \p_i-\frac{i}{2}\(\s^{\a \b}_{ij}\Th_\a \p_\b-\bar\s^{\da\db}_{ij}\bTh_{\da} \bar{\partial}_{\db}\)\,.
\ee
It is straightforward to check that the above fermionic and bosonic generators in the superspace give back the finite $\mathcal{N}=1$ CSA.

\paragraph{$\boldsymbol{R}$ symmetry generators:} Finally, we consider the superspace formulation of the $R$-symmetry generator. For $\mathcal{N}=1$, the symmetry group is $U(1)$. The relativistic $R$-symmetry generator is given by
\be{27}
R=-\frac{1}{2}(\Th^\a \p_\a-\bTh^{\dot \a}\bp_{\dot \a})\,.
\ee
Irrespective of any scaling of the $\Th$, the contracted generator remains the same, suggesting
$R\to R$. Following this scaling, the $R$ terms drop from the RHS of the $\{Q,S\}$ and $\{\bQ,\bS\}$ brackets in the algebra of section \ref{ss}. Also, the brackets with the supercharges become non-vanishing.
\be{newR}
[R,Q_\a]=-\frac{3i}{4}Q_\a\qquad\; [R,\bQ_{\da}]=\frac{3i}{4}\bQ_{\da}\qquad\; [R,S_\a]=\frac{3i}{4}S_\a\qquad\; [R,\bS_{\da}]=-\frac{3i}{4}\bS_{\da} 
\ee

\subsection*{Case 2.  Asymmetric scaling}
Here we have $a=0,\; \Th_\a \to \Th_\a, \;  \bTh_{\dot\a} \to \e\bTh_{\dot\a}$. Consider the value of $a=0$ in \eqref{avalue} and repeat the analysis similar to  Case 1 (symmetric scaling). This value of $a$ implies the following scalings for the superconformal generators:
\begin{subequations}
\label{asymm1}
\begin{align}
Q_\a &\to Q_\a & \bQ_{\da}&\to \frac{1}{\e}\bQ_{\da} & S_\a &\to \frac{1}{\e} S_\a & \bS_{\da}&\to \bS_{\da} & R&\to R\\
H&\to \frac{1}{\e}H & B_i &\to  \frac{1}{\e}B_i & K&\to  \frac{1}{\e}K & P_i&\to P_i & D&\to D & K_i&\to K_i & J_{ij}&\to J_{ij}\,.
\end{align}
\end{subequations}
It is the asymmetric scaling as discussed in Sec.~\ref{asymmetricscaling}. The superspace formulation of finite-CSA generators also justifies the scaling of the bosonic part  in \eqref{bosgen}. The algebra remains the same as \eqref{fer1}--\eqref{ral1}. Also, for this case,  
\be{29}
 \bar{Q}_{\dot \alpha}= (S_\a)^\dagger\qquad\qquad \bar{S}_{\dot \alpha}= (Q_\a)^\dagger\,.
 \ee
 There is again an identical asymmetric sector with the barred and unbarred generators interchanged. 
 
\subsection{Infinite extension of algebra} \label{infext}
In this section, we construct an infinite extension of the finite CSA \eqref{fer1}--\eqref{ral1}. We focus on $d=4$, but similar constructions work for $d>4$. We begin with the fermionic generators in the superspace coordinates. 
\begin{subequations}
\begin{align}
Q_\a&=\p_\a-\s^t_{\a\dot\b}\bar\Theta^{\db}\p_t & \bar S_{\da}&=-i \e^{\b\gamma}(\s_i)_{\gamma\da}x_i Q_\b \\
\bar Q_{\da}&=-{\bp}_{\da}+\s^t_{\b\dot\a}\Theta^{\b}\p_t & S_{\a}&=i \e^{{\db}\dot\gamma}(\s_i)_{\a\dot\gamma}x_i \bar Q_{\db}
\end{align}
\end{subequations}
Let us now define the matrix $\Lambda$, which can be used to flip from $Q$ to $\bar{S}$.
\be{lambda}
\Lambda^\b_{\;\;\;\da}=-i \e^{\b\gamma}(\s_i)_{\gamma\da}x_i\qquad\Rightarrow \qquad \bar S_{\da}=\Lambda^\b_{\;\: \da} Q_\b
\ee
Similarly, we define $\bar{\L}$ which can be used to go from $\bQ$ to $S$.
\be{barlambda}
\bar\Lambda^{\;\;\;\dot\b}_{\a}=i \e^{\db\dot\gamma}(\s_i)_{\a\dot\gamma}x_i\qquad\Rightarrow\qquad\bar S_{\a}=\bar\Lambda^{\;\;\;\db}_{\a} \bar Q_{\db}
\ee
We deduce the identities
\be{31}
\Lambda^\a_{\;\: \db}\bar\Lambda^{\;\;\dot\gamma}_{ \a}=(x_i x_i)\delta^{\dot\gamma}_{\dot\b}\qquad\qquad\Lambda^\a_{\;\: \db}\bar\Lambda^{\;\:\dot\b}_{ \gamma}=(x_i x_i)\delta^{\a}_{\gamma},\Lambda^\b_{\;\;\;\da}=[\bar\Lambda^{\;\;\;\dot\b}_{\a}]^\dagger\,.
\ee
Conventions and further identities related to $\L$ and $\bL$ are collected in Appendix \ref{identities section}. Let us now define the following fermionic generators:
\begin{subequations}
\label{fourierfermionic}
\begin{align}
G^{+}_{r_1,r_2,r_3}&=x^{r_1}y^{r_2}z^{r_3}Q_\a & G^{-}_{r_1,r_2,r_3}&=x^{r_1}y^{r_2}z^{r_3}\L_{\;\;\da}^{\a} Q_\a\\
\tG^{+}_{r_1,r_2,r_3}&=x^{r_1}y^{r_2}z^{r_3}\bar Q_{\da} & \tG^{-}_{r_1,r_2,r_3}&=x^{r_1}y^{r_2}z^{r_3}{\bar \L}_{\a}^{\;\;\da}\bar Q_{\da}\,.
\end{align}
\end{subequations}
Here, $r_i$'s can take any integer value. 
Then, 
\be{tal}
[G^{+}_{r_1,r_2,r_3},P_x]=-r_1 G^{+}_{r_1-1,r_2,r_3} \qquad\quad [G^{-}_{r_1,r_2,r_3},P_x]=-r_1 G^{-}_{r_1-1,r_2,r_3}-\p_x \L G^{+}_{r_1,r_2,r_3}\\
\ee
and similar expressions for the tilde sector with $\L$ replaced by $\bL$. 
Similarly, the fermionic generators give the following brackets with  SCT generator $K_i$:
\begin{subequations}
\label{sctal}
\begin{align}
[G^{+}_{r_1,r_2,r_3},K_x]&=(r_1+2r_2+2r_3) G^{+}_{r_1+1,r_2,r_3}-r_1\big(G^{+}_{r_1-1,r_2+2,r_3}+G^{+}_{r_1-1,r_2,r_3+2}\big)\nonumber\\
&\quad -(\L\cdot \p_x \bL )G^{+}_{r_1,r_2,r_3}\\
[G^{-}_{r_1,r_2,r_3},K_x]&=(r_1+2r_2+2r_3) G^{-}_{r_1+1,r_2,r_3}-r_1\big(G^{-}_{r_1-1,r_2+2,r_3}+G^{-}_{r_1-1,r_2,r_3+2}\big)
\end{align}
\end{subequations}
and similar expressions for the tilde sector, now with $\L\cdot \p_x \bL$ replaced by $\p_x \L \cdot \bL$. 
The brackets for the other components of $P_i$ and $K_i$ follows similarly. The brackets with the dilatation generator $D$ are given by
\be{dal}
[G^{\pm}_{r_1,r_2,,r_3},D]=(r_1+r_2+r_3 \mp \frac{1}{2}) G^{\pm}_{r_1,r_2,r_3}\,.
\ee
The same equations hold for the ${\tilde{G}}^{\pm}$.
Let us now combine both the finite and infinite fermionic generators in \eqref{fourierfermionic} to define
\be{32}
G_f=f(x_i,\L)Q_\a\,,
\ee
where  $f$ can be a function of the coordinates $x_i$ and $\L$. Thus, we have the following cases, depending on the choice of $f$:
\begin{subequations}
\label{33}
\begin{align}
 f(x_i,\L)&=1 && \Rightarrow & G_f&=Q_{\a}\\
 f(x_i,\L)&=x^{r_1}y^{r_2}z^{r_3},~\text{or } g(x_i) && \Rightarrow & G_f&=G^{+}_{r_1,r_2,r_3}\\
 f(x_i,\L)&=\L^{\a}_{\;\;\db} && \Rightarrow & G_f&=\bar S_{\db}\\
 f(x_i,\L)&=g(x_i)\L^{\a}_{\;\;\db} && \Rightarrow & G_f&=G^{-}_{r_1,r_2,r_3}\,.
\end{align}
\end{subequations}
Similarly, we define
\be{34}
\bar{G}_f=f(x_i,\bL)\bQ_{\da}
\ee
so as to combine rest of the fermionic generators in \eqref{fourierfermionic} as follows
\begin{subequations}
\label{35}
\begin{align}
f(x_i,\bL)&=1 && \Rightarrow & \bar{G}_f&=\bQ_{\da}\\
f(x_i,\bL)&=x^{r_1}y^{r_2}z^{r_3},~\text{or } g(x_i)&& \Rightarrow & \bar{G}_f&=\tilde{G}^{+}_{r_1,r_2,r_3}\\
f(x_i,\bL)&=\bL_{\b}^{\;\;\da} && \Rightarrow & \bar{G}_f&=S_{\b}\\
f(x_i,\bL)&=g(x_i)\bL_{\a}^{\;\;\db} && \Rightarrow & \bar{G}_f&=\tilde{G}^{-}_{r_1,r_2,r_3}.
\end{align}
\end{subequations}
Then, we can finally write the infinite $\mathcal{N}=1$ CSA algebra in $d=4$ succinctly as,
\be{infal1}\boxed{
\begin{aligned}
&[G_f,P_i]=G_{-\p_i f}  &[\bar{G}_f,P_i] &=\bar{G}_{-\p_i f}\\
&[G_f,D]=G_{-\frac{1}{2}f+x_i\p_i f} &[\bar{G}_f,D]&=\bar{G}_{-\frac{1}{2}f+x_i\p_i f}\\
&[G_f,K_i]=G_{-x_k x_k\p_i f+2x_i x_k\p_k f-( \L \cdot \p_i \bL)f} & [\bar{G}_f,K_i]&=\bar{G}_{-x_k x_k\p_i f+2x_i x_k\p_k f-f(\p_i \L\cdot\bL)}
\end{aligned}}
\ee
If we replace $f=x^r$ or $\L$ or $\bL$, it reproduces \eqref{tal} -- \eqref{dal}. Finally,
\be{infal2}\boxed{
\{G_f,\bar{G}_g\}=2 \s^0 M_{f(x_i,\L)\cdot g(x_i,\bL)}
}\ee
For completeness, we also present the infinite extension of the bosonic part:
\be{infalgebra}
\boxed{
\begin{aligned}
&[M_f,P_i] =M_{-\p_i f}, \quad [M_f,D] =M_{-f+x_i \p_i f}\\
& [M_f,K_i]= M_{2x_i x_k\p_k f-x_k x_k\p_i f-2x_i f}\,.
\end{aligned} }
\ee

\subsubsection*{Inclusion of $\boldsymbol{R}$ symmetry:} It is possible to give an infinite extension to the $R$-symmetry as well. Defining
\be{rsym1}
\mathfrak{R}_{r_1,r_2,r_3}=x^{r_1}y^{r_2}z^{r_3}R
\ee
yields
\begin{subequations}
\label{rsym2}
\begin{align}
&[\mathfrak{R}_{r_1,r_2,r_3},P_x]=-r_1 \mathfrak{R}_{r_1-1,r_2,r_3} \\ &[\mathfrak{R}_{r_1,r_2,,r_3},D]=(r_1+r_2+r_3) \mathfrak{R}_{r_1,r_2,,r_3} \\
&[\mathfrak{R}_{r_1,r_2,r_3},K_x]=(r_1+2r_2+2r_3) \mathfrak{R}_{r_1+1,r_2,r_3}-r_1\Big(\mathfrak{R}_{r_1-1,r_2+2,r_3}+\mathfrak{R}_{r_1-1,r_2,r_3+2}\Big)\\
&[\mathfrak{R}_{s_1,s_2,s_3},G^{\pm}_{r_1,r_2,r_3}]=-i\frac{3}{4} G^{\pm}_{r_1+s_1,r_2+s_2,r_3+s_3} \\ &[\mathfrak{R}_{s_1,s_2,s_3},\tG^{\pm}_{r_1,r_2,r_3}]=i\frac{3}{4} \tG^{\pm}_{r_1+s_1,r_2+s_2,r_3+s_3} 
\end{align}
\end{subequations}
Alternatively, \eqref{rsym1} can also be written as,
\be{36}
\mathfrak{R}_{f}=f(x_i)R
\ee
Thus, we can rewrite \eqref{rsym2} as,
\be{rsymminf}
\boxed{
\begin{aligned}
&[\mathfrak{R}_f,P_i]=\mathfrak{R}_{-\p_i f} &[\mathfrak{R}_f,D]&=\mathfrak{R}_{x_i\p_i f}\\
&[\mathfrak{R}_f,K_i]=\mathfrak{R}_{-x_k x_k\p_i f+2x_i x_k\p_k f} &&\\
&[\mathfrak{R}_g,G_f]=-i\frac{3}{4}G_{f\cdot g}  &[\mathfrak{R}_g,\bar{G}_f]&=i\frac{3}{4}\bar{G}_{f\cdot g}\\
&[\mathfrak{R}_f,M_g]=0& [\mathfrak{R}_f,\mathfrak{R}_g]&=0.
\end{aligned}}
\ee
The complete infinite $\mathcal{N}=1$  CSA in $D=4$ is given by the boxed equations \eqref{infal1}--\eqref{infalgebra} and \eqref{rsymminf}.

\section{Generalisations to higher \texorpdfstring{$\boldsymbol{\mathcal{N}}$}{N}}\label{n=4}
In the previous sections, we discussed finite and infinite  $\mathcal{N}=1$ CSA. Here, we extend our formulation for $\mathcal{N}=4$. First, let us summarise the finite superconformal generators for $\mathcal{N}=4$ CSA.  The supercharges and superconformal charges are given by, $\{Q^A_\a, \bQ^A_{\dot \a}, S^A_\a, \bS^A_{\dot \a}\}$, $(\text{where,~}A,B=1,...4)$. The $R$-symmetry group is given by $R_I$, generating an $SU(4)$ algebra.

\medskip
\noindent
Similar to $\mathcal{N}=1$, we can have two types of scalings belonging to symmetric and asymmetric cases. They are given by $ R_I \to  R_I$ in either case, and additionally
\begin{subequations}
\label{37}
\begin{align}
&\text{Symmetric:} & Q^A_\a &\to \frac{1}{ \sqrt \e} Q^A_\a & \bar Q^A_{\dot \a} &\to \frac{1}{ \sqrt\e} \bar Q^A_{\dot\a} & S^A_\a &\to \frac{1}{ \sqrt\e} S^A_\a & \bar S^A_{\dot \a} &\to \frac{1}{ \sqrt\e} \bar S^A_{\dot\a} \\
&\text{Asymmetric:} & Q^A_\a &\to \frac{1}{ \e} Q^A_\a & \bar Q^A_{\dot \a} &\to \bar Q^A_{\dot\a} & S^A_\a &\to S^A_\a & \bar S^A_{\dot \a} &\to \frac{1}{ \e} \bar S^A_{\dot\a} \,.
\end{align}
\end{subequations}
The above scalings give the following finite CSA for $\mathcal{N}=4$.

\medskip

\noindent \textit{Fermionic sector:}
\begin{subequations}
\label{n41}
\begin{align}
 \{Q^A_\a, \bar Q^B_{\dot \a} \}&=2  \sigma^0_{\a \dot \a}\delta^{AB} H & \{Q^A_\a,  Q^B_{ \b} \}&=\{\bar{Q}^A_{\dot \a},  \bar Q^B_{\dot \b} \}=0\\
 \{S^A_\a, \bar S^B_{\dot \a} \}&=2  \sigma^0_{\a \dot \a} \delta^{AB} K & \{S^A_\a,  S^B_{ \b} \}&=\{\bar{S}_{\dot \a},  \bar S_{\dot \b} \}=0\\
 \{Q^A_\a, S^B_{\b} \}&= 2 (\sigma^{0i})_\a^{~ \gamma} \e_{\gamma \b} \delta^{AB} B_i &  \{Q^A_\a, \bar S^B_{\dot \b} \}&=0 \\
 \{\bar Q^A_{\dot\a}, \bar{S}^B_{\dot\b} \}&= -2\e_{\dot \a \dot \gamma} (\bar \sigma^{0i})^{\dot \gamma}_ {~\dot \b} \delta^{AB} B_i & \{\bar Q^A_{\dot\a}, S^B_{ \b} \}&=0
\end{align}
\end{subequations}

\noindent \textit{Mixed sector:}
\begin{subequations}
\label{n42}
\begin{align}
 [ Q^A_\a,P_i]&=[Q^A_\a, H]=0 & [ S^A_\a,P_i]&=i (\sigma^i)_{\a \dot \a}\bar Q^{A\dot \a}\\
 [ \bar Q^A_{\dot\a},P_i]&=[\bar Q^A_{\dot\a}, H]=0 & [ \bar S^A_{\dot\a},P_i]&=i \e_{\dot \a \dot \b}(\bar\sigma^i)^{\dot \b \a} Q^A_{ \a}\\
 [Q^A_\a, B_i]&=[\bar Q^A_{\dot\a}, B_i]=0 & 
 [S^A_\a, H]&=[\bar S^A_{\dot \a}, H]=0\\
 [ Q^A_\a,K]&=[\bar Q^A_{\dot\a}, K]=0 & 
 [S^A_\a, B_i]&=[\bar S^A_{\dot\a}, B_i]=0\\
 [ Q^A_\a,J_{ij}]&=-\frac{i}{2}(\sigma_{ij})^\b_\a Q^A_{\b} & [ S^A_\a,K]&=[\bar S^A_{\dot\a}, K]=0 \\
 [ \bar Q^A_{\dot\a},J_{ij}]&=-\frac{i}{2}\e_{\dot \a \dot \b}(\bar \sigma_{ij})^{\dot \b}_{\dot \gamma} \bar Q^{A \dot \gamma} & 
  [ S^A_\a,K_i]&=[\bar S^A_{\dot\a}, K_i]=0\\
 [ Q^A_\a,D]&=-\frac{1}{2}Q^A_\a & 
 [ S^A_\a,J_{ij}]&=-\frac{i}{2}(\sigma_{ij})^\b_\a S^A_{\b} \\
 [ \bar Q^A_{\dot\a},D]&=-\frac{1}{2}\bar Q^A_{\dot\a} & 
 [ \bar S^A_{\dot\a},J_{ij}]&=-\frac{i}{2}\e_{\dot \a \dot \b}(\bar \sigma_{ij})^{\dot \b}_{\dot \gamma} \bar S^{A \dot \gamma}\\
 [ Q^A_\a,K_i]&=i (\sigma^i)_{\a \dot \a}\bar S^{A\dot \a} & 
 [ S^A_\a,D]&=\frac{1}{2}S^A_\a\\
  [ \bar Q^A_{\dot\a},K_i]&=i \e_{\dot \a \dot \b}(\bar\sigma^i)^{\dot \b \a} S^A_{ \a} & 
 [ \bar S^A_{\dot\a},D]&=\frac{1}{2}\bar S^A_{\dot\a}
\end{align}
\end{subequations}

\medskip
\noindent \textit{$R$-symmetry sector:}
\begin{subequations}
\label{n45}
\begin{align}
[Q^A_\a,R_I]&= \mathcal{\hat{B}}_I ^{AB}Q^B_{\a} & [\bQ^A_{\dot\a},R_I]&=-( \mathcal{\hat{B}}_I ^{AB})^\star\bQ^B_{\dot\a}\\
[S^A_\a,R_I]&=- \mathcal{\hat{B}}_I ^{AB}S^B_{\a} & [\bS^A_{\dot\a},R_I]&=( \mathcal{\hat{B}}_I ^{AB})^\star\bS^B_{\dot\a}\\
[R_I,R_J]&=i \mathfrak{t}_{IJ}^K R_K & \mathcal{\hat{B}}_I &=( \mathcal{\hat{B}}_I )^\dagger\,.
\end{align}
\end{subequations}
Using the hermiticity property of $\hat{\mathcal{B}}$ we can write $[(\hat{\mathcal{B}}_I)^{AB}]^\star=(\hat{\mathcal{B}}_I)^{BA}$. The structure constants of the $R$-symmetry are denoted by $\mathfrak{t}$. Furthermore, we can show that $\mathcal{\hat{B}}_I$ is real if we use the Jacobi identity between $\{Q,S,R_I\}$ or $\{\bar{Q},\bar{S},R_I\}$. Hence, we can write \eqref{n45} as
 \bes{}\label{n44}
\begin{align}
&[Q^A_\a,R_I]= \mathcal{\hat{B}}_I ^{AB}Q^B_{\a},~& [\bQ^A_{\dot\a},R_I]&=-( \mathcal{\hat{B}}_I ^{AB})\bQ^B_{\dot\a},\\
&[S^A_\a,R_I]=- \mathcal{\hat{B}}_I ^{AB}S^B_{\a},~& [\bS^A_{\dot\a},R_I]&=( \mathcal{\hat{B}}_I ^{AB})\bS^B_{\dot\a},\\
&[R_I,R_J]=i \mathfrak{t}_{IJ}^K R_K.
\end{align}
\ees
 The generators $\{H,B_i,K, Q^A_\a,S^A_\a, \bQ^A_{\dot \a},\bS^A_{\dot \a}\}$  are analogous to the translational part. The $R$-symmetry group is left unscaled here so that the $SU(4)$ algebra $[R_I, R_J]=i\mathfrak{t}_{IJ}^K R_K$ remains uncontracted.

 \medskip
\noindent
We can have another choice of contraction for the $R$-symmetry, $SU(4)\sim SO(6)\to ISO(5)$, such that
\be{40}
R_I \to \frac{1}{\e}\,\mathcal{R}_P, R_I
\ee
where $R_I$ has 10 generators of $SO(5)$, and $\mathcal{R}_P$ has the remaining 5 generators of  $ISO(5)$.
The resultant algebra differs from \eqref{n41},\eqref{n44}  in the following brackets:

\begin{subequations}\label{41}
\begin{align}
 \{Q^A_\a, S^B_{\b} \}&= 2 (\sigma^{0i})_\a^{~ \gamma} \e_{\gamma \b} \delta^{AB} B_i +\mathcal{\hat{B}}_P^{AB}\mathcal{R}_P,  \\
 \{\bar Q^A_{\dot\a}, \bar{S}^B_{\dot\b} \}&= -2\e_{\dot \a \dot \gamma} (\bar \sigma^{0i})^{\dot \gamma}_ {~\dot \b} \delta^{AB} B_i+(\mathcal{\hat{B}}_P^{AB})^\star \mathcal{R}_P. \\
[R_I,R_J]&=i \mathfrak{t}_{IJ}^K R_K, ~~ [\mathcal{R}_P, \mathcal{R}_S]=0, ~~[R_I,\mathcal{R}_P]=i \mathfrak{t}_{IP}^S \mathcal{R}_S\\
 [R_I, *]&=\text{ same as before},~~ [\mathcal{R}_r, *]=0. 
\end{align}
\end{subequations}

\subsection*{Infinite Extension of $\boldsymbol{\mathcal{N}=4}$ CSA in $\boldsymbol{d=4}$:}
Now, we will write down the generalisation of infinite $\mathcal{N}=4$ CSA in $d=4$. We are introducing the generalised infinite fermionic generators  $G_f^A, \bar{G}_f^A$, where $A=1,\hdots,4$.

 \medskip
\noindent
\begin{subequations}
\label{42}
\begin{align}
 f(x_i)&=1 && \Rightarrow & G_f^A &=Q_\a^A,\\
f(x_i)&=g(x_i)& &\Rightarrow & G_f^A&=g(x_i)Q_\a^A\\
f(x_i)&=\L^{\a}_{\;\;\db} && \Rightarrow & G_f^A&=\bar S_{\da}^A\\
f(x_i)&=g(x_i)\L^{\a}_{\;\;\db} && \Rightarrow & G_f^A&=g(x_i)\bar S_{\da}^A.
\end{align}
\end{subequations}

Here, $f$ can be a function of the coordinates $x_i$ and $\L$.
Also
\begin{subequations}
\begin{align}
 f(x_i)&=1 && \Rightarrow & \bar{G}_f^A &=\bQ_{\da}^A,\\
f(x_i)&=g(x_i)& &\Rightarrow & \bar{G}_f^A&=g(x_i)\bQ_{\da}^A\\
f(x_i)&=\bL_{\a}^{\;\;\db} && \Rightarrow & \bar{G}_f^A&=S_{\a}^A\\
f(x_i)&=g(x_i)\bL_{\a}^{\;\;\db} && \Rightarrow & \bar{G}_f^A&=g(x_i)\bar S_{\a}^A.
\end{align}
\end{subequations}

The resultant $\mathcal{N}=4$ infinite super CCA is
\be{infal}\boxed{
\begin{aligned}
&[G_f^A,P_i]=G_{-\p_i f}^A & [\bar{G}_f^A,P_i]&=\bar{G}_{-\p_i f}^A\\
&[G_f^A,D]=G_{-\frac{1}{2}f+x_i\p_i f}^A & [\bar{G}_f^A,D]&=\bar{G}_{-\frac{1}{2}f+x_i\p_i f}^A\\
&[G_f^A,K_i]=G_{-x_k x_k\p_i f+2x_i x_k\p_k f-f(\p_i \L\cdot\bL)}^A & [\bar{G}_f^A,K_i]&=\bar{G}_{-x_k x_k\p_i f+2x_i x_k\p_k f-f(\p_i \L\cdot\bL)}^A \\
&\{G_f^A,\bar{G}_g^B\}=2\delta^{AB} \s^0 M_{f(x_i,\L)\cdot g(x_i,\bL)}&&
\end{aligned}}
\ee

\subsubsection*{Inclusion of $\boldsymbol{R}$-symmetry:} We propose an infinite extension for the $R$-symmetry generators $R_I$.
\be{rsym3}
\mathfrak{R}^{(I)}_f=f(x^i)R_I
\ee
Then
\be{44}
\boxed{
\begin{aligned}
&[\mathfrak{R}^{(I)}_f,P_i]=\mathfrak{R}^{(I)}_{-\p_i f} & [\mathfrak{R}^{(I)}_f,D]&=\mathfrak{R}^{(I)}_{x_i\p_i f} \\
&[\mathfrak{R}^{(I)}_f,K_i]=\mathfrak{R}^{(I)}_{-x_k x_k\p_i f+2x_i x_k\p_k f} &&\\
&[G_f^A,\mathfrak{R}^{(I)}_g]=(\mathcal{\hat{B}}_I)^{AB}G^B_{f\cdot g} & [\bar{G}_f^B,\mathfrak{R}^{(I)}_g]&=-(\mathcal{\hat{B}}_I)^{AB}\bar{G}^B_{f\cdot g}~ \\
&[\mathfrak{R}^{(I)}_f,M_g]=0 &[\mathfrak{R}^{(I)}_f,\mathfrak{R}^{(J)}_g]&=i\mathfrak{t}^{K}_{IJ}\mathfrak{R}^{K}_{fg}
\end{aligned}}
\ee
Conventions and useful identities can be found in Appendix \ref{identities section}.  

\medskip

\noindent Our proposed infinite extension of the Carrollian superconformal symmetry is compatible with the known results in $d=2$ and $d=3$ boundary dimensions. We provide a detailed cross-check of the same in Appendices \ref{sec7} for $d=2$ and \ref{superbms4} for $d=3$. 

\section{Representation theory of \texorpdfstring{$\boldsymbol{\mathcal{N}=1}$}{N=1} CSA}\label{sec8}

In this section, we take the first steps towards building a representation theory for the infinite-dimensional algebras we have constructed in our previous sections. We first review relevant aspects of relativistic superconformal algebras and then focus on analogous constructions for the $\mathcal{N}=1$ CSA. Our analysis is by no means exhaustive and we plan to return to aspects of representation theory in the future. 

\subsection[Relativistic \texorpdfstring{$\mathcal{N}=1$}{N=1} chiral super multiplet]{Relativistic \texorpdfstring{$\boldsymbol{\mathcal{N}=1}$}{N=1} chiral super multiplet}\label{wz}
We begin by revisiting some relevant aspects of the representation theory of the relativistic supersymmetric algebra. We introduce first conventions here and summarise important characteristics of the Wess--Zumino chiral super multiplet structure. These discussions will help prepare the ground for the construction of the representation theory of CSA later in the section.

 \medskip
\noindent
For a generic relativistic field $\Phi(x_i,t)$ at an arbitrary spacetime point the infinitesimal supersymmetric transformation is given by
\be{45}
\de_{\a}\Phi(x_i,t)=-i[(\a Q+\bQ \bar{\a}),\Phi(x_i,t)]\,.
\ee
Here, $\a_\b,\bar{\a}_{\db}$ are relativistic anti-commuting parameters. The finite supersymmetric transformation on the generic field is written as 
\be{46}
\Phi(x_i,t)\to \Phi^\prime(x_i,t)=U^\dagger \Phi(x_i,t) U\qquad\qquad U=\exp\{-i(\a Q+\bQ \bar{\a})\}
\ee
We quote here the $\mathcal{N}=1$ relativistic algebra between the fermionic generators $Q_\a,\bQ_{\da}$.
\be{rwz}
{[\a^\b_1 Q_\b,\bQ_{\b}\bar{\a}^{\db}_2]=2\a^\b_1(\s^\mu)_{\b\db}\bar{\a}^{\db}_2P_\mu}\qquad\qquad [P_\mu,\Phi(x_i,t)]=-i\p_\mu\Phi(x_i,t)
\ee
Thus, the two successive infinitesimal SUSY transformations act on a generic field as
\be{47}
[\de_{\a_1},\de_{\a_2}]\Phi(x_i,t)=-2i(\a_1\s^\mu\bar{\a}_2-\a_2\s^\mu\bar{\a}_1)\p_\mu \Phi(x_i,t)\,.
\ee

\paragraph{Wess--Zumino multiplet:} Now, we take a complex relativistic scalar field $\phi(x_i,t)$ along with its complex conjugate $\phi^\star(x_i,t)$ to review the relativistic representation theory of $\mathcal{N}=1$ supersymmetric algebra \eqref{rwz} on a complex scalar field. The infinitesimal supersymmetric transformations $Q_\a,\bQ_{\da}$ acting on the complex scalar field  generate chiral fermions $\psi_{\a}$ (left chiral) and $\chi_{\da}$ (right chiral). In this process, the mass dimension change by $\pm \frac{1}{2}$ resulting in  chiral fermion or the derivative of the scalar field. Next, the chiral fermions of mass dimension $\frac{3}{2}$ transform into  complex auxiliary scalar field $F,F^\star$ by the action of $Q_\a,\bQ_{\da}$ on them. The mass dimension of the auxiliary field is 2. Thus, we get two sectors in the multiplet structure. The left chiral sector is generated by the following transformations
\bes{}\label{48}
\begin{align}
&\de_{\a}\phi(x)=\sqrt{2}\a\psi(x),\\
&\de_{\a}\psi(x)=\sqrt{2}i (\s^\mu\bar{\a}\p_\mu \phi(x))+\sqrt{2}\a\tilde{F}(x),\\
&\de_{\a}F(x)=-\sqrt{2}i(\p_\mu\psi(x)\s^\mu\bar{\a}).
\end{align}
\ees
The right chiral sector is generated as following
\bes{}\label{49}
\begin{align}
\de_{\a}\phi^\star(x)&=\sqrt{2}\bar{\a}\chi(x),\\
\de_{\a}\chi(x)&= -\sqrt{2}i({\a}\s^\mu\p_\mu \phi^\star(x))+\sqrt{2}\bar{\a}\tilde{F}^\star(x),\\
\de_{\a}F^\star(x)&=\sqrt{2}i(\a\s^\mu \p_\mu\chi(x)).
\end{align}
\ees
This multiplet is also known as the Wess--Zumino multiplet.

\subsection{Carrollian fermions in Weyl representation}\label{carrollian_weyl}
In this section, we discuss the Carrollian fermions in Weyl representation. The Carrollian Weyl spinors decouple and the equations of motion do not carry any interaction term. To see this, we start with the relativistic Weyl fermion
\be{50}
\Psi_{W}=\begin{bmatrix}
\psi_L\\
\chi_R
\end{bmatrix}\,.
\ee
Here, $\psi_L$ and $\chi_R$ are left and right chiral spinors. The equations of motion are
\be{51}
i\ga^\mu \p_\mu \Psi_{W}=0\qquad\im \qquad
i(\s^0\p_t\psi_L-\s^i\p_i\psi_L)=0 \qquad\qquad
 i(\s^0\p_t\chi_R+\s^i\p_i\chi_R)=0\,.
\ee
The gamma matrices are in Weyl representation
\be{52}
\gamma^0=\begin{bmatrix}
0&\s^0\\
\bar{\s}^0&0
\end{bmatrix}\qquad\qquad
\gamma^i=\begin{bmatrix}
0&\s^i\\
\bar{\s}^i&0
\end{bmatrix}\,.
\ee
The action of boosts on the chiral fermions at the origin is given by
\be{53}
[B_i,\Psi_W(0,0)]=\Sigma_{i0}\Psi_{W}(0,0)\qquad\qquad\Sigma_{i0}=-\frac{1}{4}[\ga^i,\ga^0]=\begin{bmatrix}
-\frac{\s^i}{2}&0\\
0&\frac{\s^i}{2}
\end{bmatrix}\,.
\ee
Similarly, under the rotation generators the chiral spinors transform as
\be{54}
[J_{ij},\Psi_W(0,0)]=\Sigma_{i0}\Psi_{W}(0,0)\qquad\qquad\Sigma_{ij}=-\frac{1}{4}[\ga^i,\ga^j]=\begin{bmatrix}
\frac{1}{4}[\s^i,\s^j]&0\\
0&\frac{1}{4}[\s^i,\s^j]
\end{bmatrix}\,.
\ee
Now, we are taking the UR limit on the relativistic equations of motion. Thus,
\be{55}
t\to\e t\qquad\qquad x_i\to x_i\qquad\qquad\psi_L\to \e^{r}\psi_L\qquad\qquad\chi_R\to\e^s\chi_R\qquad\qquad \e\to 0
\ee
\be{urref1}
\e^{r-1}[\s^0\p_t\psi_L-\e \s^i \p_i \psi_L]=0\qquad\qquad\e^{s-1}[\s^0\p_t\chi_R+\e \s^i \p_i \chi_R]=0\,.
\ee
Here, $r,s$ are some arbitrary parameters. We can see from equation \eqref{urref1} that the chiral spinors decouple irrespective of $r,s$ and we do not obtain any relation between $r$ and $s$. For convenience, we consider $r=s=1$ and obtain the Carrollian Weyl equations of motion from the leading order term in the $\e \to 0$ limit.
\be{56}
i\s^o\:\p_t \psi_L=0\qquad\qquad i\s^o\:\p_t \chi_R=0\,.
\ee
Thus, under rotations and boosts the spinors transform as
\bes{}\label{urref3}
\begin{align}
&[B_i,\psi_L(0,0)]=0&[B_i,\chi_R(0,0)]&=0\\
&[J_{ij},\psi_L(0,0)]=\frac{1}{4}[\s^i,\s^j]\psi_L(0,0)&[J_{ij},\chi_R(0,0)]&=\frac{1}{4}[\s^i,\s^j]\chi_R(0,0)\,.
\end{align}
\ees
Using the identity $\frac{1}{4}[\s^i,\s^j]=\frac{i}{2}(\s^{ij})=\frac{i}{2}(\bar{\s}^{ij})$, we can put back the indices to write
\bes{}\label{urref2}
\begin{align}
&[J_{ij},\psi_{\a}]=\frac{i}{2}(\s^{ij})_{\a}^{\;\;\b}\psi_{\b},\\
&[J_{ij},\chi^{\da}]=\frac{i}{2}(\bar{\s}^{ij})^{\da}_{\;\;\db}\chi^{\db},\text{and, }[J_{ij},\chi_{\da}]=\frac{i}{2}\e_{\da\db}(\bar{\s}^{ij})^{\da}_{\;\;\dot{\rho}}\e^{\dot{\rho}\dg}\chi_{\dot{\gamma}} \:.
\end{align}\ees 
We will use \eqref{urref3}, \eqref{urref2} when discussing representation theory of CSA.

\subsection[Representation theory of \texorpdfstring{$\mathcal{N}=1$}{N=1} CSA  in \texorpdfstring{$d=4$}{D=4}]{Representation theory of \texorpdfstring{$\boldsymbol{\mathcal{N}=1}$}{N=1} CSA  in \texorpdfstring{$\boldsymbol{d=4}$}{D=4}}
 To discuss the representation theory of $\mathcal{N}=1$ CSA in $d=4$, first, we consider the action of the finite set of generators on a complex scalar field and then extend our formalism to the infinite set of CSA generators. 

\medskip

\noindent We construct the highest-weight representation for $\mathcal{N}=1$ CSA on fields with different spins.  Similar to its non-supersymmetric cousin (reviewed in Sec.~\ref{Scaling for the Bosonic generators of Finite CCA}), we label the states with their scaling dimensions $\Delta$ and spin $j$. Due to $R$-symmetry transformations we have an additional label $r$ on the states. Using the state-operator correspondence, let us write
\be{59}
[D,\Phi(0,0)]=\Delta \Phi(0,0)\qquad\; [J^2, \Phi(0,0)]=j(j+1)\Phi(0,0)\qquad\; [R,\Phi(0,0)]=r\Phi(0,0)\,.
\ee
Here, $\Delta$ is the eigenvalue of the dilatation operator $D$. $J^2$ is the eigenvalue of the quadratic Casimir associated with $SO(3)$ rotations, and $j$ is the corresponding spin eigenvalue. The $R$-symmetry generator commutes with all other bosonic generators (the $R$-symmetry group for $\mathcal{N}=1$ CSA is $U(1)$). Hence, the states also carry a label $r$.

\medskip

\noindent We expect the spectrum, in particular the scaling dimension $\Delta$, to be bounded from below. This allows us to define CSA primaries in the CFT sense. Primaries are annihilated by the finite CSA generators
\be{60}
[K_i,\Phi(0,0)]=0\qquad\; [K,\Phi(0,0)]=0\qquad\; [S_\alpha,\Phi(0,0)]=0\qquad\; [\bar{S}_{\dot\alpha},\Phi(0,0)]=0\,.
\ee 

\medskip
\noindent
In the finite CSA we have $\{S,\bS\} \sim K$. Hence, the constraint regarding the temporal part of special conformal transformation $K$ $([K,\Phi(0,0)]=0)$ becomes implicit. Incorporating the infinite generators, the CSA primaries are defined as
 \bes{}\label{61}
\begin{align}
[K_i,\Phi(0,0)]&=0, &[S_\alpha,\Phi(0,0)]&=0,\;  [\bar{S}_{\dot\alpha},\Phi(0,0)]=0\\
[G^{-}_{r_1,r_2,r_3},\Phi(0,0)]&=0,&[\tG^{-}_{r_1,r_2,r_3},\Phi(0,0)]&=0~~\text{for any $r_1,r_2,r_3\geq 0$}\\
[M_f,\Phi(0,0)]&=0, &\text{for the degree of polynomial $ > 0$}.
\end{align}
\ees

\medskip
\noindent
The scaling dimensions of the fermionic operators $Q_\a,\bQ_{\da}$ are $+\frac{1}{2}$, whereas that of the fermionic superconformal operators $S_\a,\bS_{\da}$ are $-\frac{1}{2}$. The states are created by acting with $Q,\bQ$ and $P_i,H$ on the CSA primaries. Also, $Q,\bS$ have label $r=-\frac{3i}{4}$ and $S,\bQ$ have label $\frac{3i}{4}$  under the $R$ transformation.  

 \medskip
\noindent
We define the set of mutually anti-commuting parameters $\vr_\b,{\bvr}_{\db}$. Additionally, they anticommute with every fermionic object and commute with every bosonic object. For a generic field $\Phi(x_i,t)$, we define the infinitesimal transformations,
\be{62}
\delta_{\vr}\Phi(x_i,t)=[\vr^\b Q_\b,\Phi(x_i,t)]\qquad\qquad\bar{\delta}_{\bvr}\Phi(x_i,t)=[{\bQ_{\db}} \bar{\vr}^{\db},\Phi(x_i,t)]
\ee
with the convention
\be{63}
\vr Q=\vr^{\b}Q_{\b}=-\vr_{\b}Q^{\b}=Q^{\b}\vr_{\b}=Q\vr\qquad\qquad\bar{Q}\bar{\vr}=\bar{\vr}_{\db}\bar{Q}^{\db}=-\bar{Q}^{\db}\bar{\vr}_{\db}=\bar{\vr}\bar{Q}=\bar{\vr}\bQ\,.
\ee
Also, for two fermionic objects (with $A=0,i$)
\be{64}
\psi \s^A \chi=\psi^\a (\s^A)_{\a\db}\chi^{\db}\qquad\qquad \chi \bar{\s}^A \psi=\chi_{\da} (\bar\s^A)^{\da\b}\psi_\b\,.
\ee

\smallskip
\noindent

\subsection{Carrollian chiral multiplet}\label{Carrollian chiral multiplet}
\begin{figure}
\begin{center}
\includegraphics[width=0.9\linewidth]{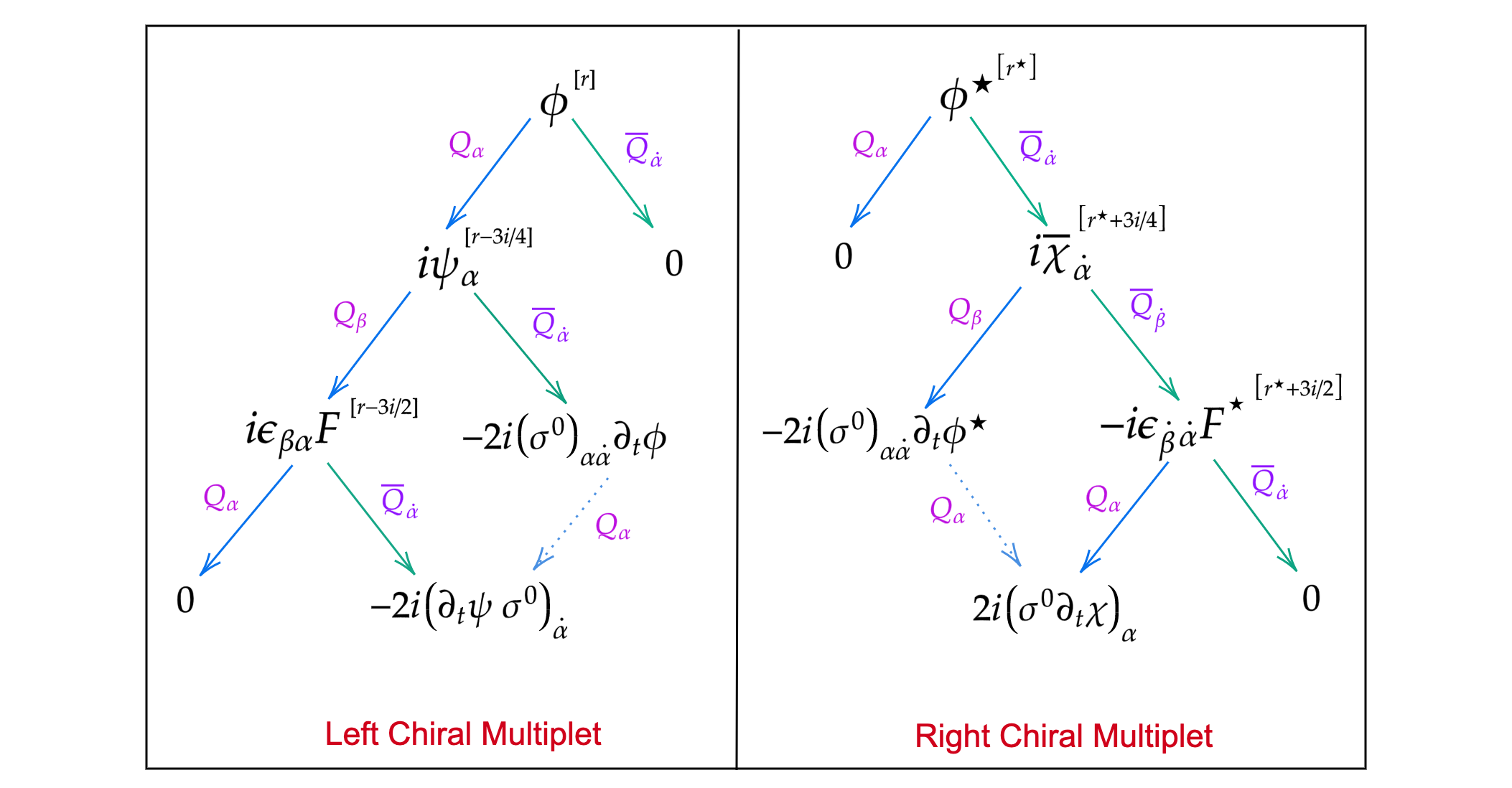}
\caption{Carrollian chiral multiplet structure}
\label{multiplet}
\end{center}
\end{figure}

In this subsection, we discuss the representation of $\mathcal{N}=1$ CSA on the fields. We start with a complex scalar field where $\Phi=\phi$ and its complex conjugate is $\phi^\star$. We consider the Carrollian fermions $\Psi$ in the Weyl representation. We also consider a complex auxiliary scalar field $F$ along with its complex conjugate $F^\star$. Let us first summarise the main results. The multiplet structure is the following:
\be{65}
\begin{split}
& \text{Complex scalars:~}  ( \phi, \phi^\star)\\
& \text{Weyl Fermions:~}  \Psi_{\text{Weyl}}=\begin{bmatrix}
\psi_{\a}\\
\chi_{\da}
\end{bmatrix}
\\ 
& \text{Complex auxiliary scalars:~}  (F, F^{\star})\,.
\end{split}
\ee
There are two distinct sectors in the multiplet structure, analogous to its relativistic Wess--Zumino counterpart. We name these two parts as left and right chiral sectors. The multiplet structures can be easily realised as displayed in Figure \ref{multiplet}.
In the left chiral sector, we have the field contents $(\phi,\psi,F)$. The fields transform in the following way under infinite $\mathcal{N}=1$ CSA generators: 
\bes{}\label{66}
\begin{align}
[G_f,\phi(x_i,t)]&=if\psi(x_i,t) &[\tilde{G}_h,\phi(x_i,t)]&=0\\ \label{eq:angelinajolie}
\{G_f,\psi(x_i,t)\}&=if\;\e
\;F(x_i,t)&\{\tilde{G}_h,\psi(x_i,t)\}&=-2i\:\tilde{h}\:\bs^0\p_t\phi(x_i,t)\\ 
[G_f,F(x_i,t)]&=0&[\tilde{G}_h,F(x_i,t)]&=-2i\;\tilde{h}\;(\p_t\psi(x_i,t)\s^0)\\
[\mathfrak{R}_{k},\phi(x_i,t)]&=r\:\hat{k}\:\phi(x_i,t)&
[\mathfrak{R}_{k},\psi(x_i,t)]&=(r-\frac{3i}{4})\:\hat{k}\:\psi(x_i,t)\\
[\mathfrak{R}_{k},F(x_i,t)]&=(r-\frac{3i}{2})\:\hat{k}\:F(x_i,t)
\end{align}
\ees
where
\be{67}
G_f=f(x_i,\L)Q\qquad\qquad \tilde{G}_h=\tilde{h}(x_i,\bL)\bQ\qquad\qquad\mathfrak{R}_{k}=\hat{k}(x_i)R\,.
\ee
Just to explain the notation, we remark that the left anti-commutator \eqref{eq:angelinajolie} for $f=1$ expands as $\{Q_\alpha,\psi_\beta(x_i,t)\}=i\;\e_{\b\a}\;F(x_i,t)$ in component notation.

 \medskip
\noindent
Similarly, for the right chiral sector, we have $(\phi^\star,\chi,F^\star)$. Under infinite $\mathcal{N}=1$ super Carrollian conformal generators the right chiral sector emerges as
\bes{}\label{68}
\begin{align}
[G_f,\phi^\star(x_i,t)]&=0&[\tilde{G}_h,\phi^\star(x_i,t)]&=i\;\tilde{h}\;\chi(x_i,t)\\
\{G_f,\chi(x_i,t)\}&=-2i\:f\:\s^0\p_t\phi^\star(x_i,t)&\{\tilde{G}_h,\chi(x_i,t)\}&=-i\;\tilde{h}\;\e
\;F^\star(x_i,t)\\ 
[G_f,F^\star(x_i,t)]&=2i\;f\;(\s^0\p_t\chi(x_i,t))&[\tilde{G}_h,F^\star(x_i,t)]&=0\\
[\mathfrak{R}_{k},\phi^\star(x_i,t)]&=r^\star\:\hat{k}\:\phi^\star(x_i,t)&\\
[\mathfrak{R}_{k},\chi(x_i,t)]&=(r^\star+\frac{3i}{4})\:\hat{k}\:\chi(x_i,t)& [\mathfrak{R}_{k},F^\star(x_i,t)]&=(r^\star+\frac{3i}{2})\:\hat{k} F^\star(x_i,t)\,.
\end{align}
\ees 
For further details of the representation theory we recall the action of Carrollian boost and Carrollian rotation generators on fields of different spins. We start with a generic complex scalar primary $\phi$ at the origin. The rotation and boost generators operate on a scalar as 
\be{69}
[J_{ij},\phi(0,0)]=0\qquad\qquad[B_i,\phi(0,0)]=0\,.
\ee
For Carrollian Weyl spinors, we can write
\bes{}\label{70}
\begin{align}
[J_{ij},\psi_{\a}(0,0)]&=\frac{1}{4}[\s_i,\s_j]_{\a}^{\:\:\b}\psi_{\b}(0,0)=\frac{i}{2}(\s_{ij})_{\a}^{\:\:\b}\psi_{\b}(0,0)\\
[J_{ij},\chi_{\da}(0,0)]&=\frac{1}{4}\e_{\da\db}[\s_i,\s_j]^{\db}_{\:\:\dot{\rho}}\e^{\dr{\dg}}\chi_{\dg}(0,0)=\frac{i}{2}\e_{{\da}\db}(\bar{\s}_{ij})^{\db}_{\:\:\dot{\rho}}\e^{{\dr}\dg}\chi_{\dg}(0,0)\\
[B_i,\psi_\a(0,0)]&=[B_i,\chi_{\da}(0,0)]=0\,.
\end{align}
\ees

\subsubsection*{Left chiral sector} We are now equipped to understand the transformation of fields under different fermionic generators. We  use the Jacobi identities to find the infinitesimal transformations. The details of the calculations are presented in the Appendix \ref{representation1}.
\bes{}\label{529}
\begin{align}
\{B_i,\phi(0,0),Q_{\a}\}:&\im  [Q_\a,\phi(0,0)]=i\psi_\a(0,0)\qquad\delta_{\vr}\phi(0,0)=i\vr\psi(0,0)\\\label{boost1}
\{B_i,\phi(0,0),\bQ_{\da}\}:&\im [[\bQ_{\da},\phi],B_i]=0\\
&\im [\bQ_{\da},\phi(0,0)]=0\qquad \bar{\de}_{\bvr} \phi(0,0)=0\\
\{J_{ij},\phi(0,0),Q_{\a}\}: &\implies [J_{ij},\psi_{\a}(0,0)]=\frac{i}{2}(\s_{ij})_{\a}^{\:\:\b}\psi_{\b}(0,0)\\
\{Q_{\a},\bQ_{\da},\phi(0,0)\}:& \im \{\bQ_{\da},\psi_{\a}\}=-2i(\s^0)_{\a\da}\p_t\phi(0,0)\\
&\im \bar{\delta}_{\bvr}\psi_{\b}(0,0)=2i(\s^0 \bvr)_{\b}\p_t \phi(0,0)\\
\{J_{ij},Q_\a,\psi_\b(0,0)\}: &\im \{Q_\a,\psi_\b(0,0)\}=i\e_{\b\a}F(0,0)\\
&\im\de_{\vr}\psi_\b (0,0)=i\vr_\b F(0,0)\\
\{Q_{\a},\bQ_{\da},\psi_\b(0,0)\}:&\im [\bQ_{\da},F(0,0)]=-2i(\p_t\psi\;\s^0)_{\da}\\
&\im\bar{\de}_{\bvr} F(0,0)=-2i(\p_t\psi\;\s^0\;\bvr)\\
\{Q_{\a},\bQ_{\da},F(0,0)\}:&\im [Q_\a,F(0,0)]=0\qquad\de_{\vr}F(0,0)=0
\end{align}
\ees
The transformations above are consistent with the Jacobi identities involving all other finite CSA generators. We have also checked for consistency of the above infinitesimal transformations with our finite CSA. The details of the calculations are provided in Appendix  \ref{representation1}. 

\noindent The super Carrolian conformal primaries also carry $R$ charges. Let
\be{73}
[R,\phi]=r\phi\,.
\ee
Then, we can further show from the Jacobis
\bes{}\label{73.5}
\begin{align}
& \{Q_\a,R,\phi\}: &\im&& [R,\psi_\a]&=(r-\frac{3i}{4})\psi_\a\\
& \{Q_\a,\psi_\b,R\}: &\im&& [R,F]&=(r-\frac{3i}{2})F\,.
\end{align}
\ees

Next we consider the action of other fermionic generators $S_\a,\bS_{\da}$ on a CSA primary at a generic spacetime point. For a CSA primary $\Phi$ 
\be{74}
[S_\a,\Phi(0,0)]=[{\bS}_{\da},\Phi(0,0)]=0\,.
\ee
Also, the transformation of  primaries at a generic spacetime point can be obtained as
\be{75}
[\mathcal{O},\Phi(x_i,t)]=U[U^{-1}\mathcal{O}U,\Phi(0,0)]U^{-1}\qquad\qquad U=e^{-tH-x_iP_i}
\ee
where $\mathcal{O}$ denotes a generic CSA generator. Using the Baker--Campbell--Hausdorff formula, we can evaluate the transformation rules of the primaries at a generic spacetime point under the fermionic generators 
\bes{}\label{leftS1}
\begin{align}
[S_\a,\phi(t,x_i)]&=0\\
[\bS_{\da},\phi(t,x_i)]&=\e_{\da\db}(x_i \bar{\sigma}_i)^{\db\a}\:\psi_{\a}(t,x_i)\\
\{S_\a,\psi_{\b}(t,x_i)\}&=-2(x_i \sigma_i)_{\a\da}\:\e^{\da\db}(\bar{\s}^0)_{\db\b}\p_t\phi(t,x_i)\\
\{{\bS}_{\da},\psi_{\b}(t,x_i)\}&=-\e_{\da\db}(x_i \bar{\sigma}_i)^{\db\a}\e_{\a\b}F(t,x_i),\\
[{\bS}_{\db},F(t,x_i)]&=0\\
[S_\b,F(t,x_i)]&=-2(x_i \sigma_i)_{\b\db}\:\e^{\db\dg}(\bar{\s}^0)_{\dg\rho}\p_t\psi^{\rho}(t,x_i)\,.
\end{align}
\ees
The details of these derivations are provided in Appendix \ref{representation1}.

\medskip

Finally, we consider the action of infinite-dimensional fermionic generators on the fields. These infinite-dimensional fermionic generators can be written as
\be{76}
G_f=f(x_i,\L)Q\qquad\qquad \tilde{G}_h=\tilde{h}(x_i,\bL)\bQ\qquad\qquad\mathfrak{R}_{k}=\hat{k}(x_i)R\,.
\ee
The primaries of different spins transform as
\bes{}\label{77}
\begin{align}
[G_f,\phi(x_i,t)]&=if\psi(x_i,t),~~&[\tilde{G}_h,\phi(x_i,t)]&=0\\
\{G_f,\psi(x_i,t)\}&=if\;\e_{L}\;F(x_i,t),~~&\{\tilde{G}_h,\psi(x_i,t)\}&=-2i\:\tilde{h}\:\bs^0\p_t\phi(x_i,t),~~\e_L=\e_{\b\a}\\
[G_f,F(x_i,t)]&=0,~~&[\tilde{G}_h,F(x_i,t)]&=-2i\;\tilde{h}\;(\p_t\psi(x_i,t)\s^0)\\
[\mathfrak{R}_{k},\phi(x_i,t)]&=r\:\hat{k}\:\phi(x_i,t)\\
[\mathfrak{R}_{k},\psi(x_i,t)]&=(r-\frac{3i}{4})\:\hat{k}\:\psi(x_i,t)\\
[\mathfrak{R}_{k},F(x_i,t)]&=(r-\frac{3i}{2})\:\hat{k}\:F(x_i,t)\,.
\end{align}
\ees
As a consistency check, notice that putting $f=\L,\tilde{h}=\bL$ in the above equations reproduces the commutators \eqref{leftS1}.

\paragraph{Left chiral sector from limit:} So far, we have studied various aspects of the representation theory of $\mathcal{N}=1$ CSA in $d=4$ purely from an intrinsic approach which involves studying the action of various generators acting on the CSA primaries. Here, we examine the consistency of our construction taking the limiting approach. This involves taking the UR limit on the parent relativistic representation theory of $\mathcal{N}=1$ supersymmetric theory. We restrict ourselves only on the Wess--Zumino multiplet reviewed in section \ref{wz}.
The relativistic infinitesimal transformations of the Wess--Zumino multiplets are given by 
\bes{}\label{relleft1}
\begin{align}
&\de_{\a}\phi(x)=\sqrt{2}\a\psi(x)\\
&\de_{\a}\psi(x)=\sqrt{2}i (\s^\mu\bar{\a}\p_\mu \phi(x))+\sqrt{2}\a\tilde{F}(x)\\
&\de_{\a}F(x)=-\sqrt{2}i(\p_\mu\psi(x)\s^\mu\bar{\a})\\
\non &\text{where, }\de_{\a}=-i(\a Q+\bQ \bar{\a}),~\a_\b,\bar{\a}_{\db}: ~~\text{relativistic anti-commuting parameters.}
\end{align}
\ees
Take the asymmetric scaling \eqref{asymm1},
\be{78}
Q\to Q\qquad\qquad\bQ\to \frac{1}{\e}\bQ\qquad\implies\qquad \de_{\a}\to -i(\de_\vr+\frac{1}{\e}\bde_{\bvr})
\ee
together with the scaling of the complex scalar field and Weyl chiral fermions, along with the usual spacetime scalings
\be{79}
\phi \to \frac{1}{\e}\phi\qquad\qquad\psi\to \frac{1}{\e}\psi\qquad\qquad F\to \frac{1}{\e}F\qquad\qquad \p_t\to\frac{1}{\e}\p_t\qquad\qquad \p_i \to \p_i\,.
\ee
Inserting these scalings in \eqref{relleft1}, and equating the orders of $\frac{1}{\e}$ on both sides of the infinitesimal transformations yields
\bes{}\label{80}
\begin{align}
-i\de_{\vr}\phi&=\sqrt{2}\vr \psi,&\bde_{\bvr}\phi&=0,\\
-i\bde_{\bvr}\psi&=\sqrt{2}i((\s^0\bvr)\p_t\phi+\e(\s^i\bvr) \p_i\phi), &i\de_{\vr}\psi&=\sqrt{2}\vr F,\\
\de_\vr F&=0,&i\bde_{\bvr}F&=-\sqrt{2}i((\p_t\psi \:\s^0\:\bvr)+\e(\p_i\psi \:\s^i\:\bvr)).
\end{align}
\ees
Finally, in the limit $\e\to0$,
\be{81}\boxed{
\phantom{\Bigg(}
\begin{aligned}
\de_{\vr}\phi&=i\sqrt{2}\vr \psi&\bde_{\bvr}\phi&=0\\
\de_{\vr}\psi&=\sqrt{2}i\vr F &\bde_{\bvr}\psi&=-\sqrt{2}(\s^0\bvr)\p_t\phi\\
\de_\vr F&=0&\bde_{\bvr}F&=\sqrt{2}(\p_t\psi \:\s^0\:\bvr)\,.
\end{aligned}
\phantom{\Bigg)}}
\ee
These infinitesimal transformations are consistent with the transformation rules obtained in an intrinsic analysis without the normalization constants.

\medskip
\noindent
\subsection*{Right chiral sector} Finally, we focus on the right chiral sector of the $\mathcal{N}=1$ CSA representation theory. We repeat an analysis similar to its left sided counterpart, without presenting all intermediate steps, since the calculations are analogous to the left chiral counterpart. The details of the calculations for the Jacobi identities are described in Appendix \ref{representation2}. Here, we only mention the most important results, especially when they differ from the left chiral sector. At $t=x_i=0$ we have the (graded) commutation relations
\bes{}
\label{540}
\begin{align}
\label{540a} [\bQ_{\da},\phi^\star]&=i\chi_{\da}& [Q_{\a},\phi^\star]&=[\bQ_{\da},F^{\star}]=0 \\ \label{540b}
\{Q_{\a},\chi_{\da}\}&=-2i(\s^0)_{\a\da}\p_t\phi^\star& \{\bQ_{\da},\chi_{\db}\}&=-i\e_{\db{\da}}F^\star\\ \label{540c}
[Q_{\a},F^\star]&=2i(\s^0)_{\a\da}\e^{\da\dr}\p_t\chi_{\dr}  &[J_{ij},\chi_{\da}]&=\frac{i}{2}\e_{\da\db}(\bar{\s}^{ij})^{\db}_{\:\:\dot{\rho}}\e^{\dot{\rho}\dg}\chi_{\dg}\\ \label{544}
[R,\phi^{\star}]=r^\star\phi^{\star}\quad [R,\chi_{\da}]&=(r^\star+\frac{3i}{4})\chi_{\da} & [R,F^{\star}]&=(r+\frac{3i}{2})F^{\star}\,.
\end{align}
\ees

\noindent The $S$-transformations give
\bes{}\label{rightS}
\begin{align}
[S_\a,\phi^\star(t,x_i)]&=(x_i {\sigma}_i)_{\a\da}\e^{\da\db}\:\chi_{\db}(t,x_i)&
[\bS_{\da},\phi^\star(t,x_i)]&=0\\
\nonumber \{{\bS}_{\da},\chi_{\db}(t,x_i)\}&=-2\e_{\da\db}(x_i \bar{\sigma}_i)^{\db\a}\:({\s}^0)_{\a\dr}\p_t\phi^\star(t,x_i)&
\{S_\a,\chi_{\db}(t,x_i)\}&=-(x_i {\sigma}_i)_{\a\db}F^\star(t,x_i)\\
[{\bS}_{\da},F^\star(t,x_i)]&=2\e_{\da\db}(x_i \bar{\sigma}_i)^{\db\a}\:({\s}^0)_{\a\dr}\p_t\chi^{\dr}(t,x_i)&
[S_\a,F^\star(t,x_i)]&=0\,.
\end{align}
\ees
Thus, we can find the transformation rules of the right chiral sector under $\mathcal{N}=1$ finite CSA generators, consistent with the $\mathcal{N}=1$ CSA algebra. Finally, we write the transformation rules for the infinite (fermionic) symmetry generators resembling \eqref{68}.
\noindent
As a check, we can put $f=\L,\tilde{h}=\bL$ in the equations \eqref{68}, which reproduce the $S$-transformation relations \eqref{rightS}.\\
 \medskip
\noindent
\paragraph{Right chiral sector from the limit:} After the intrinsic analysis, we revisit the right chiral sector from the limiting point of view. The relativistic infinitesimal transformations of the right chiral multiplet are given as
\bes{}\label{relright}
\begin{align}
\de_{\a}\phi^\star(x)&=\sqrt{2}\bar{\a}\chi(x),&
\de_{\a}\chi(x)&= -\sqrt{2}i({\a}\s^\mu\p_\mu \phi^\star(x))+\sqrt{2}\bar{\a}\tilde{F}^\star(x),\\
\de_{\a}F^\star(x)&=\sqrt{2}i(\a\s^\mu \p_\mu\chi(x)),\\
\non\text{where }\de_{\a}&=-i(\a Q+\bQ \bar{\a}),&\a_\b,\bar{\a}_{\db} &\text{: relativistic anti-commuting parameters.}
\end{align}
\ees

 \medskip
\noindent
We take the asymmetric scaling \eqref{assym3} ($Q\to \frac{1}{\e}Q, \: \bQ\to\bQ$) along with the scaling of the complex scalar field and Weyl chiral fermions, and the usual spacetime scaling,  implying
\be{84}
\de_{\a}\to-i(\frac{1}{\e}\de_\vr+\bde_{\bvr}),~~ \phi^\star \to \frac{1}{\e}\phi^\star,~~ \chi\to \frac{1}{\e}\chi,~~ F^\star\to \frac{1}{\e}F^\star,~~ \p_t\to\frac{1}{\e}\p_t,~~ \p_i \to \p_i\,.
\ee
Plugging these scalings into \eqref{relright} and equating the orders $\frac{1}{\e}$ on both sides of the infinitesimal transformations, obtains the following transformation relations in the limit $\e\to0$
\be{87}
\boxed{
\phantom{\Bigg(}
\begin{aligned}
\de_{\vr}\phi^\star&=0&\bde_{\bvr}\phi^\star&=i\sqrt{2}\bvr \chi\\
\de_{\vr}\chi&=\sqrt{2}(\vr\s^0)\p_t\phi^\star&\bde_{\bvr}\chi&=i\sqrt{2}\bvr F^\star\\
\de_{\vr}F^\star&=-\sqrt{2}(\vr\s^0\p_t\chi)&\bde_{\bvr} F^\star&=0\,.
\end{aligned}
\phantom{\Bigg)}
}
\ee
These infinitesimal transformations are analogous to the left chiral counterpart \eqref{81} and are consistent with the transformation rules we obtained in intrinsic analysis without the normalization constants.

\section{Conclusions}
\label{Conclusions}

We have explored the algebraic structures associated with Carrollian superconformal symmetry --- or equivalently super BMS symmetry --- in the context of field theories in dimensions greater than three. We specifically concentrated on boundary dimension $d=4$, but the method outlined is simply generalisable for all higher dimensions. We identified the contraction of the corresponding relativistic algebra that is useful for holography of asymptotically flat spacetimes and then lifted the finite contracted algebra to an infinite-dimensional one. Our constructions generalised to higher $\mathcal{N}$. We checked our proposed infinite lift by reproducing results in the literature for $d=2, 3$. Finally, we returned to $\mathcal{N}=1$ Carrollian superconformal field theories and worked out details of its representation theory. In particular, we discussed the Carrollian version of the Wess--Zumino multiplet, laying the groundwork for similar constructions for other multiplets to be addressed in the future. 

\subsection{Discussion and future directions}
The algebraic construction in this paper is relevant for understanding Carrollian superconformal field theories. Our final aim is to build a Carrollian version of $\mathcal{N}=4$ $SU(N)$ super Yang--Mills theory so that we have a concrete proposal for the boundary theory in the flat limit of Maldacena's original AdS/CFT correspondence. Building on the representation theory of the $\mathcal{N}=1$ Carrollian superconformal field theories in the present paper, we aim to construct explicit examples of Carrollian superconformal field theories, starting with $\mathcal{N}=1$ Carrollian SUSY electrodynamics, before moving to Yang--Mills and higher $\mathcal{N}$. It could be rewarding to find all possible central extensions of our algebras \eqref{infal1}-\eqref{infalgebra} and \eqref{infal}. Our preliminary analysis for finding central terms by educated guessing was unsuccessful, but we did not attempt a complete analysis of finding non-trivial co-cycles.

\medskip

\noindent The fact that the underlying symmetry structures are infinite-dimensional may indicate some hidden integrability in the Carrollian sector of these SUSY gauge theories. $\mathcal{N}=4$ SYM provides a natural playground to investigate integrability \cite{Beisert:2010jr}. A first step would be to understand what exists in the planar limit of the Carrollian $\mathcal{N}=4$ SYM, which possibly can be obtained directly from the corresponding UR limit. But there could be more structure in the Carrollian sector and perhaps integrability beyond the planar limit. If this were true, we would have stumbled on a new integrable sector of relativistic $\mathcal{N}=4$ SYM, viz., the Carrollian sector.

\medskip

\noindent When considering the flat version of the AdS$_5$/CFT$_4$ correspondence, it is necessary to understand the 'tHooft large-$N$ limit in the Carrollian context. Here the speed of light goes to zero along with the usual 'tHooft scaling of 
$N\to \infty$ and fixed $\lambda\equiv g_{\text{YM}}^2 N$.
There is a myriad of possibilities on how to take this modified large $N$ limit, generalising the original analysis in relativistic gauge theories. Now that we understand the algebraic structures that govern the symmetries of the Carrollian superconformal theory we are after, the possible large $N$ limits have to be consistent with the particular scalings discussed in our paper. It could be useful to first understand this in the $\mathcal{N}=1$ context before generalising to the case of interest, viz., $\mathcal{N}=4$.   

\medskip

\noindent After the large $N$ limit is established, the next step would be to construct the bulk-boundary dictionary between flat space supergravity (and, ultimately, superstrings) and these Carrollian superconformal theories. For the AdS$_5$/CFT$_4$ correspondence, the curvature radii of the AdS$_5$ and the accompanying S$^5$ are equal and it would seem that a flat limit on the AdS$_5$ needs to be accompanied by a similar flat limit on the S$^5$, thus making the resulting theory a 10-dimensional flat space theory. This should be reflected in terms of a contracted $R$-symmetry of the form we discussed in our generalisations to $\mathcal{N}=4$ in Sec.~\ref{n=4}. Eventually, we intend to understand the bulk-boundary dictionary both as a limit of the original correspondence and intrinsically. It is likely that the limit only yields a part of the answer, as the contraction of symmetries generates only the Poincar{\'e} subalgebra of the BMS algebra and not the entire symmetry structure. Our expectation is that the infinite-dimensional symmetries constructed in our work will make it easier to understand holography in flat space in general dimensions.

\bigskip

\paragraph{Note added:} While completing this work a paper was posted on the {\tt arXiv}, Ref.~\cite{Banerjee:2022abf}, that has overlap with ours. While their focus is on boundary dimensions $d=2$ and $d=3$, our focus is on boundary dimension $d>3$, specifically $d=4$.

\bigskip

\section*{Acknowledgements}
It is a pleasure to thank Stephane Detournay, Hong Liu, and Joan Sim\'on for discussions at early stages, Matthias Heumesser, and Wout Merbis at intermediate stages, and Rudranil Basu at all stages of this project.

\smallskip 

 \noindent AB is partially supported by a Swarnajayanti fellowship of the Department of Science and Technology, India and by the following grants from the Science and Engineering Research Board: SB/SJF/2019-20/08, MTR/2017/000740, CGR/2020/002035. 
 
 \noindent DG was supported by the Austrian Science Fund (FWF), projects P~30822, P~32581 and P~33789.
 
\noindent PN was supported by the Science and Engineering Research Board (SERB-OVDF) fellowship ODF/2018/000759 during the early stages of this work.


\appendix
\section{Relativistic \texorpdfstring{$\boldsymbol{SU(2,2|N)}$}{SU(2,2|N)} algebra}\label{relal}
Let us first summarise the relativistic superconformal $SU(2,2|N)$ algebra here. \\
\textit{Fermionic sector:}
\begin{subequations}
\begin{align}
 \{Q^A_\a, \bar Q^B_{\dot \a} \}&=2 \delta^{AB} \sigma^\mu_{\a \dot \a}P_\mu\qquad\qquad \{Q^A_\a,  Q^B_{ \b} \}=\{\bar{Q}^A_{\dot \a},  \bar Q^B_{\dot \b} \}=0\\
 \{S^A_\a, \bar S^B_{\dot \a} \}&=2 \delta^{AB} \sigma^\mu_{\a \dot \a}K_\mu\qquad\qquad \{S^A_\a,  S^B_{ \b} \}=\{\bar{S}^A_{\dot \a},  \bar S^B_{\dot \b} \}=0\\
 \{Q^A_\a, S^B_{\b} \}&= 2 \e_{\a \b}\delta^{AB} D-2 (\sigma^{\mu \nu})_\a^{~ \gamma} \e_{\gamma \b}J_{\mu \nu} \delta^{AB}
-4i \e_{\a \b}\delta^{AB}R+ \mathcal{\hat{B}}_i ^{AB}R_i\\
 \{\bar Q^A_{\dot\a}, \bar{S}^B_{\dot\b} \}&= 2 \e_{\dot\a \dot\b}\delta^{AB} D+2\e_{\dot \a \dot \gamma} (\bar \sigma^{\mu \nu})^{\dot \gamma}_ {~\dot \b}J_{\mu \nu} \delta^{AB}+4i \e_{\dot\a \dot\b}\delta^{AB}R+( \mathcal{\hat{B}}_i ^{AB})^\star R_i\\
\{Q^A_\a, \bar S^B_{\dot \b} \}&=\{\bar Q^A_{\dot\a}, S^B_{ \b} \}=0
\end{align}
\end{subequations}

\noindent\textit{Mixed sector:}
\begin{subequations}
\begin{align}
  [ Q^A_\a,P_\mu]&=0 &  [ S^A_\a,P_\mu]&=i \sigma^\mu_{\a \dot\a} \bar Q^{A\dot\a}\\
   [ \bar Q^A_{\dot\a},P_\mu]&=0& [ \bar S^A_{\dot \a},P_\mu]&=i\e_{\dot \a\dot \b}( \bar{\sigma}^{\mu })^{\dot\b \a}  Q^{A}_\a \\
  [ Q^A_\a,K_\mu]&=i \sigma^\mu_{\a \dot\a} \bar S^{A\dot\a} & [ S^A_\a,K_\mu]&=0\\
  [ \bar Q^A_{\dot\a},K_\mu]&=i\e_{\dot \a\dot \b}( \bar{\sigma}^{\mu })^{\dot\b \a}  S^{A}_\a& [ \bar S^A_{\dot\a},K_\mu]&=0\\
  [ Q^A_\a,J_{\mu \nu}]&=-\frac{i}{2}(\sigma_{\mu\nu})^\b_\a Q^A_{\b} & [ S^A_\a,J_{\mu \nu}]&=-\frac{i}{2}(\sigma_{\mu\nu})^\b_\a S^A_{\b}\\
  [ \bar Q^A_{\dot\a},J_{\mu \nu}]&=-\frac{i}{2}\e_{\dot \a \dot \b}(\bar \sigma_{\mu\nu})^{\dot \b}_{\dot \gamma} \bar Q^{A \dot \gamma}& [ \bar S^A_{\dot\a},J_{\mu \nu}]&=-\frac{i}{2}\e_{\dot \a \dot \b}(\bar \sigma_{\mu\nu})^{\dot \b}_{\dot \gamma} \bar S^{A \dot \gamma}\\
  [ Q^A_\a,D]&=-\frac{1}{2}Q^A_\a& [ S^A_\a,D]&=\frac{1}{2}S^A_\a,\\
 [ \bar Q^A_{\dot\a},D]&=-\frac{1}{2}\bar Q^A_{\dot\a}& [ \bar S^A_{\dot\a},D]&=\frac{1}{2}\bar S^A_{\dot\a}
\end{align}
\end{subequations}

\noindent\textit{$R$-symmetry sector:}
\begin{subequations}
\begin{align}
&[Q^A_\a, R_i]= \mathcal{\hat{B}}_i ^{AB}Q^B_{\a},~~[\bar Q^A_{\dot \a}, R_i]=-( \mathcal{\hat{B}}_i ^{AB})^\star \bar Q^B_{\dot \a},~~[S^A_\a, R_i]=- \mathcal{\hat{B}}_i ^{AB}S^B_{\a}, ~~[\bar S^A_{\dot \a}, R_i]=( \mathcal{\hat{B}}_i ^{AB})^\star \bar S^B_{\dot \a}\nonumber\\
&[R,Q^A_\a]=-i \Big(\frac{4-N}{N}\Big)Q^A_\a,~~[R,\bar Q^A_{\dot \a}]=i \Big(\frac{4-N}{N} \Big) \bar Q^A_{\dot \a},~~[R,S^A_\a]=i \Big(\frac{4-N}{N}\Big)S^A_{\a},\\
&[R,\bar S^A_{\dot \a}]=-i \Big(\frac{4-N}{N} \Big) \bar S^A_{\dot \a},~~[R_i, R_j]=i \mathfrak{t}_{ij}^{ k}R_k, ~~[\text{Poincar{\'e}},R]=[\text{Poincar{\'e}},R_i]=0
\end{align}
\end{subequations}
For $N=1$, the $R$-symmetry group is $U(1)$ and $R_i=0$, whereas for $N=4$, the $R$-symmetry group is $SU(4)$ and $R=0$. $ \mathcal{\hat{B}}_i $ is a hermitean matrix. 


\section{Isomorphism with super-BMS algebra}\label{appenb}

In this appendix, we discuss our claim in \eqref{CB}. We show that the proposed isomorphism for the supersymmetric extension of the BMS algebra works in lower dimensions, $d=2,3$. We reviewed two different types of supersymmetric extension of BMS algebra in section \ref{sec:3.1}. Here, we describe higher $\mathcal{N}$ extensions of the homogeneous super BMS algebra.

\subsection*{Homogeneous super BMS with $\boldsymbol{R}$-symmetry and higher $\boldsymbol{\mathcal{N}}$}
 The homogeneous super BMS$_3$ algebras described in section 3.1 are devoid of any internal symmetry.  The authors in \cite{Caroca:2018obf} constructed the $\mathcal{N}=2$ super BMS$_3$ algebra starting from $\mathcal{N}=(2,0)$ AdS$_3$ supergravity. However, they did not use the algebraic contraction of the super Virasoro algebra. Instead, they constructed the $\mathcal{N}$ extended homogeneous super BMS$_3$ algebra from the semi-group expansion method (S-expansion) from only one copy of the higher $\mathcal{N}$ super Virasoro algebra.  The inclusion of $R$-symmetry is reflected in the $\mathcal{N}=2$ super BMS$_3$ algebra as
\begin{subequations}
\label{withR}
\begin{align}
[L_m,\tau_n]&=-n\tau_{m+n} & [M_m,\tau_n]&=-n \mathcal{Z}_{m+n}\\
[L_m, \mathcal{Z}_n]&=-n \mathcal{Z}_{m+n} & [\tau_m,\tau_n]&=\frac{c_L}{3}m\delta_{m+n,0}\\
[\tau_m,\mathcal{Z}_n]&=\frac{c_M}{3}m\delta_{m+n,0} & [Q^\a_r,\tau_m]&=\e^{\a\b}Q^\b_{m+r}\\
\{Q^{\a}_r,Q^{\b}_s\}&=\delta^{\a \b}\left[M_{r+s}+\frac{c_M}{6}\(r^2-\frac{1}{4}\)\delta_{r+s,0}\right]&&\hspace*{-2.4truecm}+2 \e^{\a\b}(r-s)\mathcal{Z}_{r+s}\,. 
\end{align}
\end{subequations}
The rest of the brackets in \eqref{noR} are also present in the $\mathcal{N}=2$ SBMS$_3$ algebra \eqref{withR}. The same algebra \eqref{noR}--\eqref{withR} can also be obtained as a limit of $\mathcal{N}=(2,2)$ AdS$_3$ supergravity (truncating the fermionic generators in one of the copies) \cite{Fuentealba:2017fck}. A similar construction can be carried out for higher supersymmetric extension of the BMS$_3$ algebra \cite{Banerjee:2018hbl}.\\
\smallskip
\noindent
 We briefly mention here the $\mathcal{N}=4$ super BMS$_3$ algebra obtained from the S-expanded $\mathcal{N}=4$ super Virasoro algebra \cite{Caroca:2018obf}.
\begin{subequations}
\label{3.4}
\begin{align}
[L_m,L_n]&=(m-n)L_{m+n} & [L_m,M_n]&=(m-n)M_{m+n}\\
[L_m,G^{I\pm}_{-\frac{1}{2}+r}]&=(\frac{m+1}{2}-r)G^{I\pm}_{-\frac{1}{2}+m+r} & \{G^{I+}_{-\frac{1}{2}+r},G^{J-}_{-\frac{1}{2}+s}\}&=\delta^{IJ}M_{r+s-1}\\
[L_m,\tau^a_{n}]&=-n\tau^a_{m+n} & [M_m,\tau^a_{n}]&=0\\
[\tau^a_m,G^{I+}_{-\frac{1}{2}+r}]&=-\frac{1}{2}(\s^a)^{IJ}G^{J+}_{m+r-\frac{1}{2}} & [\tau^a_{m},\tau^b_{n}]&=i\epsilon^{abc}\tau^c_{m+n}\\
[\tau^a_m,G^{I-}_{-\frac{1}{2}+r}]&=\frac{1}{2}(\tilde{\s}^a)^{IJ}G^{J-}_{m+r-\frac{1}{2}} & \text{where\,}(\tilde{\s}^a)_{IJ}&=-(\s^a)_{JI}\,.
\end{align}
\end{subequations}

\subsection[Boundary dimension \texorpdfstring{$d=2$}{d=2}]{Boundary dimension \texorpdfstring{$\boldsymbol{d=2}$}{d=2}}\label{sec7}
As an exercise, we construct the $\mathcal{N}=1$ CSA in $d=2$ from our construction of $\mathcal{N}=1$ finite CSA \eqref{fer1}--\eqref{ral1}. To read the CSA$_2$, we are considering only one spatial dimension $i=x_1$. We do not have the Carrollian rotation generators $J_{ij}$ in the 2d algebra. 
Let us explore the isomorphism between  CSA$_2$  and super BMS$_3$ algebras by considering the combination
\begin{subequations}
\label{comd=2}
\begin{align}
Q^{+}_{-\frac{1}{2}}&=\frac{1}{2}(Q+{\bar Q}) & Q^{-}_{+\frac{1}{2}}&=\frac{1}{2}(S+\bS)\\
Q^{+}_{+\frac{1}{2}}&=\frac{1}{2}(\p_i{\bar\L}\bS+\p_i \L S) & Q^{-}_{-\frac{1}{2}}&=\frac{1}{2}(\p_i{\L}Q+\p_i \bar\L \bQ)\,.
\end{align}
\end{subequations}
Here, we are defining \be{d=2ss}
\bS=\L Q\qquad\qquad S={\bar\L}\bQ.
\ee 
The matrices $\L,\bL$ are given by
\be{lbll}
\L=\begin{bmatrix}
-ix_1 & 0\\
0&ix_1
\end{bmatrix}\qquad\qquad
\bar{\L}=\begin{bmatrix}
ix_1& 0\\
0&-ix_1
\end{bmatrix}\,.
\ee
Here, equations \eqref{d=2ss} and \eqref{lbll} are the $d=2$ analogue of \eqref{lambda}, \eqref{barlambda}.
We give the combinations of the finite generators \eqref{comd=2} an infinite lift in $d=2$ 
\be{match1d=2}
Q^{+}_{-\frac{1}{2}+r}=\frac{1}{2}x_1^r(Q+{\bar Q})\qquad\qquad Q^{-}_{-\frac{1}{2}+r}=\frac{1}{2}x_1^r(\p_i \L Q+\p_i {\bar\L} \bQ)\,.
\ee
Here, $r$ takes arbitrary integer values. We used the identity
$\L\cdot \p_i {\bar\L}={\bar\L}\cdot \p_i\L=x_1$.
We also have the bosonic generators  $L_{(-1,0,1)}=-P_i,D,K_i$ and $M_{(-1,0,1)}=H,-B_i,K$ respectively for global part of $BMS_3$. Using \eqref{match1d=2}, we can rewrite the infinite $\mathcal{N}=1$ CSA in $d=2$ as
\begin{subequations}
\label{n=1match}
\begin{align}
[L_n,L_m]&=(n-m)L_{n+m} &
[L_n,M_m]&=(n-m)M_{n+m}\\
[L_n,Q^{\a}_r]&=(\frac{n}{2}-r)Q^{\a}_{n+r} &
\{Q^{\a}_r,Q^{\b}_s\}&=\delta^{\a \b}M_{r+s}\qquad \a,\b=\pm\,. 
\end{align}
\end{subequations}
Here, we have used the scaling $R\to \e^n R\:\: (n>0)$ as discussed after \eqref{rdiffscal}. This particular choice of scaling removes the internal symmetry from the algebra.
The algebra \eqref{n=1match} is the homogeneous super BMS$_3$ algebra as described in \eqref{noR} with vanishing central charges. It is also isomorphic to the homogeneous super Galilean Conformal algebra with vanishing central charges.

\smallskip 
\noindent \textit{Inclusion of $R$-symmetry:} 
Now let us identify 
\be{90}
\tau_0=-\frac{4}{3}\s_3 R
\ee
which we give it an infinite extension
\be{91}
\tau_n=-\frac{4}{3}\s_3 x_1^n R\,.
\ee
Here, $\s_3$ is the diagonal Pauli matrix and $n$ takes integer values. We can write the algebra containing the $R$-symmetry part as
\begin{subequations}
\label{n=1matchR}
\begin{align}
[L_m,\tau_n]&=-n\tau_{m+n} &
[M_m,\tau_n]&=0 &
[\tau_m,\tau_n]&=0\\
[Q^\a_r,\tau_m]&=\e^{\a\b}Q^\b_{m+r} & \e^{+-}&=1 & \e^{-+}&=-1\,.
\end{align}
\end{subequations}
The algebra \eqref{n=1matchR} along with the mixed sector \eqref{n=1match} matches with the homogeneous super BMS$_3$ algebra \eqref{withR} including the $R$-symmetry, with vanishing central charges, $c_1=c_2=\mathcal{Z}_r=0$.

\subsection*{Higher $\boldsymbol{\mathcal{N}}$}\label{n=4match}
Now, we justify the scaling in \eqref{37}, which resulted in the finite $\mathcal{N}=4$ CSA \eqref{n41}--\eqref{n44}. We reduce the algebra to $d=2$ by considering only one spatial direction $i=x$.  Next, we match the higher $\mathcal{N}$ formulation of super CCA in $d=2$ with known homogeneous super BMS$_3$ algebra.

\smallskip
\noindent
To see this, let us take the $\mathcal{N}=2$ formulation of CSA in $d=2$, where we have only one spatial direction. 
Putting $A,B=1,2$ in \eqref{n41}--\eqref{n44}, we obtain $\mathcal{N}=2$ CSA with the supercharges $Q^1,~ Q^2, ~ S^1, ~ S^2,~
\bQ^1,~ \bQ^2, ~ \bS^1, ~ \bS^2$ and $R$-symmetry generators $R_a\in SU(2), ~R \in U(1)$. Let us identify the combinations for the fermionic generators
\begin{subequations}
\label{n=2match}
\begin{align}
G^{1+}_{-\frac{1}{2}}&=\frac{1}{2}(Q^1+\bQ^2) &
G^{1+}_{+\frac{1}{2}}&=\frac{1}{2}(\p_i\bL \bS^1+\p_i \L\bS^2)=\frac{x}{2}(Q^1+\bQ^2)\\
G^{1-}_{-\frac{1}{2}}&=\frac{1}{2}(\bQ^1+Q^2) &
G^{1-}_{+\frac{1}{2}}&=\frac{1}{2}(\p_i\L S^1+\p_i \bL\bS^2)=\frac{x}{2}(\bQ^1+Q^2)\\
G^{2+}_{-\frac{1}{2}}&=\frac{1}{2}(Q^2-\bQ^1) &
G^{2+}_{+\frac{1}{2}}&=\frac{1}{2}(\p_i\bL \bS^2-\p_i \L S^1)=\frac{x}{2}(Q^2-\bQ^1)\\
G^{2-}_{-\frac{1}{2}}&=\frac{1}{2}(\bQ^2-Q^1) &
G^{2-}_{+\frac{1}{2}}&=\frac{1}{2}(\p_i\L S^2-\p_i \bL\bS^1)=\frac{x}{2}(\bQ^2-Q^1)\,.\\
\hat{\mathcal{B}}_a&=-\frac{1}{2}\s_a &
\mathfrak{t}_{ab}^{~c}&=\e^{abc}\;\text{(upper and lower indices are the same)}.
\end{align}
\end{subequations}
We have used the identities mentioned in \eqref{d=2ss} and \eqref{lbll} in the combinations above. We have also scaled the $U(1)$ charge as $R\to \e^n R~(n >0) $.
We give the identifications in \eqref{n=2match} an infinite lift,
\be{infn=2match}
G^{I\pm}_{-\frac{1}{2}+r}=x^r G^{I\pm}_{-\frac{1}{2}}~~~~~I=1,2\qquad\qquad
\tau^a_m=x^m R_a~~~~~m,r\in\mathbb{Z}\,.
\ee
The finite generators in \eqref{n=2match} and subsequently their infinite extensions in \eqref{infn=2match} give the algebra
\begin{subequations}
\label{95}
\begin{align}
[L_m,L_n]&=(m-n)L_{m+n}\\
[L_m,M_n]&=(m-n)M_{m+n}\\
[L_m,G^{I\pm}_{-\frac{1}{2}+r}]&=(\frac{m+1}{2}-r)G^{I\pm}_{-\frac{1}{2}+m+r}\\
\{G^{I+}_{-\frac{1}{2}+r},G^{J-}_{-\frac{1}{2}+s}\}&=\delta^{IJ}M_{r+s-1}\\
[L_m,\tau^a_{n}]&=-n\tau^a_{m+n}\\
[M_m,\tau^a_{n}]&=0\\
[\tau^a_m,G^{I+}_{-\frac{1}{2}+r}]&=-\frac{1}{2}(\s^a)^{IJ}G^{J+}_{m+r-\frac{1}{2}}\\
[\tau^a_m,G^{I-}_{-\frac{1}{2}+r}]&=\frac{1}{2}(\tilde{\s}^a)^{IJ}G^{J-}_{m+r-\frac{1}{2}}\qquad\qquad(\tilde{\s}^a)_{IJ}=-(\s^a)_{JI},\\
[\tau^a_{m},\tau^b_{n}]&=i\epsilon^{abc}\tau^c_{m+n}\,.
\end{align}
\end{subequations}
 The algebra \eqref{95} matches with the higher $\mathcal{N}$ homogeneous super BMS$_3$ algebra in \eqref{3.4}, with central charges $c_1=c_2=\mathcal{Z}_r=0$. 

\subsection[Boundary dimension \texorpdfstring{$d=3$}{d=3}]{Boundary dimension \texorpdfstring{$\boldsymbol{d=3}$}{d=3}}\label{superbms4}
In this appendix, we compare the infinite CSA in $d=3$ with the known results of super BMS$_4$. First, we concentrate on the finite CSA \eqref{fer1}--\eqref{mix11}. We reduce the constructed algebra to $d=3$, choosing $i=2$.

\medskip
\noindent
The BMS$_4$ algebra can be realised from CCA in $d=3$ with the following identifications. For the bosonic generators of the CCA in $d=3$, we can write \cite{Bagchi:2012cy}
\begin{subequations}
\label{finbms4}
\begin{align}
L_0&=\frac{1}{2}(D+i J_{xy}) & L_{-1}&=\frac{i}{2}(P_x+iP_y) & L_1&=\frac{i}{2}(K_x-iK_y)\\
\bar{L}_0&=\frac{1}{2}(D-i J_{xy}) & \bar{L}_{-1}&=\frac{i}{2}(P_x-iP_y) & \bar{L}_1&=\frac{i}{2}(K_x+iK_y)\\
M_{\frac{1}{2},\frac{1}{2}}&=-K\quad\; M_{-\frac{1}{2},-\frac{1}{2}}=H & M_{-\frac{1}{2},\frac{1}{2}}&=-i(B_x+i B_y)& M_{\frac{1}{2},-\frac{1}{2}}&=-i(B_x-i B_y)\,.
\end{align}
\end{subequations}
The generators of f-CCA in $d=3$ form the BMS$_4$ algebra for $n=0,\pm 1$ \cite{Bagchi:2012cy,Fotopoulos_2020}.
\begin{subequations}
 \label{96}
 \begin{align}
 [L_n,L_m]&=(n-m)L_{n+m} & [\bar{L}_n,\bar{L}_m]&=(n-m)\bar{L}_{n+m}&[M_{r,s},M_{p,q}]&=0\\
[L_n,M_{r,s}]&=\(\frac{n}{2}-r\)M_{n+r,s}&[\bar{L}_n,M_{r,s}]&=\(\frac{n}{2}-s\)M_{r,n+s}\,. && 
 \end{align}
\end{subequations}
 The generators in \eqref{finbms4} can be given at infinite lift for all integer values of $n,r,s$ \cite{Bagchi:2012cy}. 
 \begin{subequations}
 \label{d.3}
 \begin{align}
L_n&=-\frac{1}{2}(ix+y)^{n}[(n+1)\p_t+(x-iy)(\p_x+i\p_y)]\\
\bar{L}_n&=-\frac{1}{2}(ix-y)^{n}[(n+1)\p_t+(x+iy)(\p_x+i\p_y)]\\
M_{-\frac{1}{2}+r,-\frac{1}{2}+s}&=(ix+y)^r(ix-y)^s\p_t
 \end{align}
 \end{subequations}
Thus, the CCA in $d=3$ is isomorphic to the BMS algebra in $d=4$. Next, we focus on the fermionic generators of the $\mathcal{N}=1$ CSA in $d=3$. We write the fermionic generators of the finite-CSA as
\be{d.4}
G_{-\frac{1}{2}}=\frac{i}{2}(\bQ_{\dot{1}}+Q_2),~G_{\frac{1}{2}}=\frac{i}{2}(S_1+\bS_{\dot{2}}),~\bar{G}_{-\frac{1}{2}}=-\frac{i}{2}(Q_1+\bQ_{\dot{2}}),~\bar{G}_{\frac{1}{2}}=\frac{i}{2}(\bS_{\dot{1}}+S_2)\,.
\ee
We use the identities \eqref{lambda}, \eqref{barlambda} to give the finite generators in \eqref{d.4} an infinite lift
\be{d5}
G_{-\frac{1}{2}+r}=(ix+y)^r G_{-\frac{1}{2}}\qquad\qquad\bar{G}_{-\frac{1}{2}+r}=(ix-y)^r \bar{G}_{-\frac{1}{2}}\,.
\ee
Here, $r=1$ give back the fermionic generators of finite-CSA in $d=3$. However, the generators can be given an infinite lift for arbitrary integer values of $r$. We scale the $R$-symmetry as $R\to \e R$. The bosonic generators in \eqref{d.3} together with the fermionic generators in \eqref{d5} construct the super-BMS$_4$ algebra \cite{Fotopoulos:2020bqj}.
\be{sbms4}
[L_n,G_k]=\(\frac{n}{2}-k\)G_{n+k}\qquad\quad[\bar{L}_n,\bar{G}_r]=\(\frac{n}{2}-r\)\bar{G}_{n+r}\qquad\quad\{G_m,\bar{G}_n\}=M_{m,n}
\ee
We have only written the non-vanishing commutators of super-BMS$_4$. However,
we can construct another copy of \eqref{sbms4} by writing the fermionic generators of CSA$_3$ as
\be{97}
\begin{split}
N_{-\frac{1}{2}}=\frac{i}{2}(\bQ_{\dot{1}}-Q_2),~N_{\frac{1}{2}}=\frac{i}{2}(S_1-\bS_{\dot{2}}),~\bar{N}_{-\frac{1}{2}}=-\frac{i}{2}(Q_1-\bQ_{\dot{2}}),~\bar{N}_{\frac{1}{2}}=\frac{i}{2}(\bS_{\dot{1}}-S_2)\,.
\end{split}
\ee
The generators can be given an infinite lift for all integer values of $r$.
\be{98}
N_{-\frac{1}{2}+r}=(ix+y)^r N_{-\frac{1}{2}},~~\bar{N}_{-\frac{1}{2}+r}=(ix-y)^r \bar{N}_{-\frac{1}{2}}\,.
\ee
We consider the $R$-symmetry scaling $R \to \e R$ similar to the previous case.
Finally, we can write the super BMS$_4$ which is isomorphic to CSA$_3$
\be{99}
[L_n,N_k]=\(\frac{n}{2}-k\)N_{n+k}\qquad\quad[\bar{L}_n,\bar{N}_r]=\(\frac{n}{2}-r\)\bar{N}_{n+r}\qquad\quad\{N_m,\bar{N}_n\}=M_{m,n}\,.
\ee
The commutators between $\{G_{m}$ or $\bar{G}_{n}\}$ and $\{N_{m}$ or $\bar{N}_{n}\}$ vanish with the scaling $R \to \e R$.

\section{Identities}\label{identities section}
In this appendix, we list identities used in sections \ref{sec3}--\ref{sec8}. In all formulas below upper and lower indices for $\s$ matrices are the same. In particular, we use the following formula collection:
\begin{subequations}
\label{100}
\begin{align}
 (\s^0)_{\a\db}&=\begin{bmatrix}
1 & 0\\
0&1
\end{bmatrix} &
(\s^1)_{\a\db}&=\begin{bmatrix}
0 & 1\\
1&0
\end{bmatrix} & 
(\s^2)_{\a\db}&=\begin{bmatrix}
0 & -i\\
i&0
\end{bmatrix} &
(\s^3)_{\a\db}&=\begin{bmatrix}
1 & 0\\
0&-1
\end{bmatrix}\\
\s^{01}&=\begin{bmatrix}
0 & -i\\
-i&0
\end{bmatrix} &
\s^{02}&=\begin{bmatrix}
0 & -1\\
1&0
\end{bmatrix} &
\s^{03}&=\begin{bmatrix}
-i & 0\\
0&i
\end{bmatrix} & 
\s^{12}&=\begin{bmatrix}
1 & 0\\
0&-1
\end{bmatrix}\\
\s^{13}&=\begin{bmatrix}
0 & i\\
-i&0
\end{bmatrix} &
\s^{23}&=\begin{bmatrix}
0 & 1\\
1&0
\end{bmatrix} &
\bar\s^{01}&=\begin{bmatrix}
0 & i\\
i&0
\end{bmatrix} &
\bar\s^{02}&=\begin{bmatrix}
0 & 1\\
-1&0
\end{bmatrix} \\
\bar\s^{03}&=\begin{bmatrix}
i & 0\\
0&-i
\end{bmatrix} & \bar\s^{12}&=\begin{bmatrix}
1 & 0\\
0&-1
\end{bmatrix} & \bar\s^{13}&=\begin{bmatrix}
0 & i\\
-i & 0
\end{bmatrix} & \bar\s^{23}& =\begin{bmatrix}
0 & 1\\
1&0
\end{bmatrix}
\end{align}
\end{subequations}
Besides the explicit matrix representations above, we also use the abstract formulas
\begin{subequations}
\begin{align}
(\bar{\s^0})^{\da \b}&=({\s^0})_{\a \db} & (\bar{\s^i})^{\da \b}&=-({\s^i})_{\a \db}\\
(\bar\s^\mu)^{\da \a}&=\e^{\da \db}\e^{\a \b}(\s_\mu)_{\b \db} & (\s_\mu)_{\a \da}&=\e_{\a \b}\e_{\da \db}(\bar\s^\mu)^{\db \b} \\
(\s^{\mu \nu})^{\a\b}&=\frac{i}{2}(\s^\mu \bar\s^\nu-\s^\nu \bar\s^\mu)^{\a\b} & (\s^{\mu \nu})_{\a\b}&=\frac{i}{2}(\s^\mu \bar\s^\nu-\s^\nu \bar\s^\mu)_{\a\b}\\
(\bar\s^{\mu \nu})^{\da\db}&=\frac{i}{2}(\bar\s^\mu \s^\nu-\bar\s^\nu \s^\mu)^{\da\db} & (\bar\s^{\mu \nu})_{\da\db}&=\frac{i}{2}(\bar\s^\mu \s^\nu-\bar\s^\nu \s^\mu)_{\da\db}
\end{align}
\end{subequations}
and
\begin{subequations}
\begin{align}
(\s^{\mu \nu})_{\a}^{\;\;\b}&=\e_{\a\rho}(\s^{\mu\nu})^{\rho \b}=\frac{i}{2}(\s^\mu \bar\s^\nu-\s^\nu \bar\s^\mu)_{\a}^{\;\;\b}\\
(\bar\s^{\mu \nu})^{\da}_{\;\;\db}&=(\bar\s^{\mu \nu})^{\dot{\a}\dot{\rho}}\e_{\db\dot{\rho}}=-(\bar\s^{\mu \nu})^{\dot{\a}\dot{\rho}}\e_{\dot{\rho}\db}=\frac{i}{2}(\bar\s^\mu \s^\nu-\bar\s^\nu \s^\mu)^{\da}_{\;\;\db}=[(\s^{\mu \nu})_{\a}^{\;\;\b}]^\dagger\\
(\s^{\mu \nu})_{\a}^{\;\;\b}\e_{\b \s}&=(\s^{\mu \nu})_{\s}^{\;\;\b}\e_{\b \a}\,.
\end{align}
\end{subequations}
Our sign conventions for the $\e$-symbols are fixed as
\begin{align}
\e^{\a\b}&=\e^{\da \db}=\begin{bmatrix}
0 & 1\\
-1&0
\end{bmatrix} & \e_{\a\b}&=\e_{\da \db}=\begin{bmatrix}
0 & -1\\
1&0
\end{bmatrix}\,.
\end{align}

\subsection*{Identities used for $\boldsymbol{d=4}$}
Some identities involving $\L$ and $\bar{\L}$ in $d=4$ are
\bes{}\label{1033}
\begin{align}
\L&=\begin{bmatrix}
-ix_1+x_2 & ix_3\\
ix_3&ix_1+x_2
\end{bmatrix} &
\bar{\L}&=\begin{bmatrix}
ix_1+x_2 & -ix_3\\
-ix_3&-ix_1+x_2
\end{bmatrix}\\
\p_j \Lambda&=-i \e^{\b\ga} (\s_j)_{\ga\da} & \p_j \bar{\Lambda}&=i \e^{\db\dg} (\s_j)_{\a\da}=-i (\s_j)_{\a\dg}\e^{\dg\db}\\
\p_1 \Lambda&=-i \s_3\qquad \p_2 \Lambda=\s_0\qquad\p_3 \Lambda=i \s_1 & \p_1 \bar\Lambda&=i \s_3\qquad \p_2 \Lambda=\s_0\qquad\p_3 \Lambda=-i \s_1
\end{align}
\ees
and
\bes{}\label{104}
\begin{align}
\Lambda^\b_{\;\;\;\da}&=-i \e^{\b\gamma}(\s_i)_{\gamma\da}x_i&&\e^{\b \ga}(\s \cdot x)_{\ga \da}=[\e^{\db\dot\gamma}(\s \cdot x)_{\a\dot\gamma}]^\dagger\\
\bar\Lambda^{\;\;\;\dot\b}_{\a}&=i \e^{\db\dot\gamma}(\s_i)_{\a\dot\gamma}x_i&&(\s^0)_{\a\db}\L^{\b}_{\;\;\db} H=-\e_{\da\dg}(\bar\s^{0i}\cdot B_i)^{\dg}_{\;\;\db}\\
 \Lambda^\b_{\;\;\;\da}&=[\bar\Lambda^{\;\;\;\dot\b}_{\a}]^\dagger&&(\s \cdot x)_{\a\dg} \e^{\dg \db}(\s^0)_{ \db \b}H=i(\s^{0i}\cdot B_i)_{\a}^{\;\;\ga}\e_{\ga \b}\\
 \p_i\Lambda^\b_{\;\;\;\da}&=[\p_i\bar\Lambda^{\;\;\;\dot\b}_{\a}]^\dagger&&(\s^{0i}\cdot B_i)_{\a}^{\;\;\ga}\e_{\ga \b}= {\bL}_{\a}^{\;\;\db}(\s^0)_{\db \b}H\\
\Lambda^\a_{\;\: \db}\bar\Lambda^{\;\;\dot\gamma}_{ \a}&=(x_i x_i)\delta^{\dot\gamma}_{\dot\b},&&(\p_i \L \cdot \bL)\cdot \L= x_k x_k \p_i \L\\
\Lambda^\a_{\;\: \db}\bar\Lambda^{\;\:\dot\b}_{ \gamma}&=(x_i x_i)\delta^{\a}_{\gamma} &&(\p_i \bL)_{\a}^{\;\;\db}=-i(\s^i)_{\a\da}\e^{\da \db}\\
\p_i \Lambda^\a_{\;\;\da}&=-i\e^{\a\b}(\s^i)_{\b\dot\a}=-i\e_{\dot\a \dot\b}({\bar\s}^i)^{\dot\b \a}&&(\p_i \L)_{\da}^{\;\;\a}=-i\e_{\da \db}(\bar\s^i)^{\db\a}\\
x_k \p_k \L&=\L,~ x_k \p_k \bL=\bL&& \L\cdot \p_i \bL+\p_i\L\cdot\bL=2x_i\,.
\end{align}
\ees

\subsection*{Identities used for $\boldsymbol{d=2}$}
Here, we list identities involving $\L,\bar{\L}$ in $d=2$. We have only one spatial coordinate $i=x$ and one time coordinate.
\bes{}\label{105}
\begin{align}
-\p_i \L \cdot \s^{0i} \cdot \e &=\p_i {\bar\L} \cdot {\dot\e} \cdot {\bar\s}^{0i}=\mathbb{I}_2, &&
\p_i \L \cdot \p_i {\bar\L}= \p_i {\bar\L} \p_i \L =\mathbb{I}_2\\
(\p_i \L)^2&= (\p_i {\bar\L})^2=-\mathbb{I}_2,&&
\L\cdot \p_i {\bar\L}={\bar\L}\cdot \p_i \L=x \mathbb{I}_2\\
\L \cdot \bar\L&= \bar\L\cdot \L=x^2 \mathbb{I}_2,&&
\p_i \L=\begin{bmatrix}
i & 0\\
0&-i
\end{bmatrix},~ \p_i \bar\L=\begin{bmatrix}
-i & 0\\
0&i
\end{bmatrix}
\end{align}
\ees
Many of the relations in \eqref{1033}-\eqref{104} simplify as $\L,\bar\L$ are all diagonal in $d=2$.

\section{Details of representation theory}\label{appen3}
This appendix is intended to complement the calculations in section \ref{Carrollian chiral multiplet}. We discuss details of the Carrollian chiral multiplet and the transformation relations of the fields in the multiplet under $\mathcal{N}=1$ infinite Carrollian superconformal generators. For convenience, we rewrite here the field contents of the Carrollian chiral multiplet in equation \eqref{65}
\be{}
\begin{split}
&\text{Complex scalars:~}( \phi, \phi^\star)\\
&\text{Weyl Fermions:~}\Psi_{\text{Weyl}}=\begin{bmatrix}
\psi_{\a}\\
\chi_{\da}
\end{bmatrix}\\
& \text{Complex auxiliary scalars:~} (F, F^{\star}).
\end{split}
\ee

We have divided this appendix into two subsections, focusing on the left and right chiral sectors, respectively.

\subsection{Left chiral sector}\label{representation1}
Here, we give details of the calculations used in the representation theory of $\mathcal{N}=1$ Carrollian superconformal theories. More specifically, we show below the details of the transformation relations in \eqref{529}, \eqref{73.5} and \eqref{leftS1}. We start with calculating the Jacobi identities mentioned in \eqref{529}. The following Jacobi identities capture the transformation relations of the fields in the theory under fermionic generators.
\be{leftphi1}
\begin{aligned}
&\bullet\{B_i,\phi(0,0),Q_{\a}\}:\\
&\implies {[[B_i,\phi],Q_\a]}+[[\phi,Q_\a],B_i]+[[Q_\a,B_i],\phi]=0,\\
&\implies {[[Q_\a,\phi],B_i]}=0,\\
&\implies\boxed{ [Q_\a,\phi(0,0)]=i\psi_\a(0,0).}\\
&\text{Also,~} [\vr^\a Q_\a,\phi(0,0)]=i\vr^{\a}\psi_\a(0,0),\\
&\implies\boxed{\delta_{\vr}\phi(0,0)=i\vr\psi(0,0).}~~~~~~~\text{(where,~}\vr\psi=\vr^{\a}\psi_{\a}=-\vr_{\a}\psi^{\a},~\delta_{\vr}=\vr^\a Q_\a=-\vr_\a Q^\a).
\end{aligned}
\ee
\be{leftphi2}
\hspace{-7cm}
\begin{aligned}
\noindent&\bullet\{B_i,\phi(0,0),\bQ_{\da}\}:\\
&\implies {[[\bQ_{\da},\phi],B_i]}=0,\implies \boxed{[\bQ_{\da},\phi(0,0)]=0.}\\
&\text{Also,~}  [\bQ_{\da}{\bvr}^{\da},\phi(0,0)]=0,\implies \boxed{\bar{\de}_{\bvr} \phi(0,0)=0.}
\end{aligned}
\ee

\be{106}
\hspace{-6cm}
\begin{aligned}
&\bullet\{J_{ij},\phi(0,0),Q_{\a}\}:\\
&\implies{[[J_{ij},\phi],Q_{\a}]}+[[\phi,Q_\a],J_{ij}]+[[Q_\a,J_{ij}],\phi]=0,\\
&\implies -\frac{i}{2}(\s^{ij})_{\a}^{\:\:\b}[Q_{\b},\phi]+[[\phi,Q_\a],J_{ij}]=0,\\
&\implies[J_{ij},-i\psi_{\a}(0,0)]=-\frac{i}{2}(\s_{ij})_{\a}^{\:\:\b}[Q_{\b},\phi(0,0)]\\
&\implies\boxed{[J_{ij},\psi_{\a}(0,0)]=\frac{i}{2}(\s_{ij})_{\a}^{\:\:\b}\psi_{\b}(0,0).}
\end{aligned}
\ee

\be{leftpsi1}
\begin{aligned}
&\bullet \{Q_{\a},\bQ_{\da},\phi(0,0)\}:\\
&\implies [\{Q_{\a},\bQ_{\da}\},\phi]+\{[\phi,Q_{\a}],\bQ_{\da}\}+\{[\phi,\bQ_{\da}],Q_{\a}\}=0,\\
&\implies{[2(\s^0)_{\a\da}H,\phi]}+\{-i\psi_\a,\bQ_{\da}\}=0,\\
&\implies\boxed{\{\bQ_{\da},\psi_{\a}\}=-2i(\s^0)_{\a\da}\p_t\phi(0,0).}\\
&\text{Also,~} [ \bvr^{\da}\bQ_{\da},\psi_{\b}(0,0)]=-2i\bvr^{\da}(\s^0)_{\b\da}\p_t\phi(0,0),~~~~~~\text{(using,~}\bvr^{\da}Q_{\da}=-\bvr_{\da}Q^{\da}=-\bar{\de}_{\bvr})\\
&\implies\boxed{\bar{\delta}_{\bvr}\psi_{\b}(0,0)=2i(\s^0 \bvr)_{\b}\p_t \phi(0,0).}~~~~~~~~~~~~~~~~~~\text{(where,~}(\s^0 \bvr)_{\b}=(\s^0)_{\b\da}\bvr^{\da})
\end{aligned}
\ee

\bea{d.5}
\non&&  \hspace{-3cm} \bullet  \{J_{ij},Q_\a,\psi_\b(0,0)\}: \\
\non&&  \hspace{-3cm} \im [\{Q_\a,\psi_\b\},J_{ij}]+\{[J_{ij},\psi_{\b}],Q_{\a}\}+\{[J_{ij},Q_{\a}],\psi_\b\}=0,\\
\non&&  \hspace{-3cm} \im [J_{ij},\{Q_\a,\psi_\b(0,0)\}]=\{[J_{ij},\psi_\b],Q_\a\}+\{[J_{ij},Q_\a],\psi_\b\},\\
&&   \hspace{-3cm} ~~~~~~~~~~~~~~~~~~~~~~~~~~~= \frac{i}{2}\left((\s_{ij})_\b^{\;\;\rho}\{\psi_\rho(0,0),Q_\a\}+(\s_{ij})_\a^{\;\;\rho}\{Q_\rho,\psi_\b(0,0)\} \right)
\eea
The LHS of \eqref{d.5} should be zero, as $\{Q_a,\psi_{\b}\}$ gives a scalar. Let us write,
\be{107}
\{Q_\a,\psi_\b(0,0)\}=i\e_{\b\a}F(0,0).
\ee
Then the RHS of \eqref{d.5} gives,
\be{leftpsi2}
\begin{aligned}
&\frac{i}{2}\left((\s_{ij})_\b^{\;\;\rho}\e_{\rho\a}+(\s_{ij})_\a^{\;\;\rho}\e_{\b\rho} \right)F(0,0)\\
&=\frac{i}{2}\left((\s_{ij})_\b^{\;\;\rho}\e_{\rho\a}-(\s_{ij})_\a^{\;\;\rho}\e_{\rho\b} \right)F(0,0)\\
&=0.     ~~~~~~~~~~ \text{(Using the identity:} (\s_{ij})_\b^{\;\;\rho}\e_{\rho\a}=(\s_{ij})_\a^{\;\;\rho}\e_{\rho\b})
\end{aligned}
\ee
Hence,
\be{leftF1}
 \hspace{-4cm}
\begin{aligned}
&\boxed{\{Q_\a,\psi_b(0,0)\}=i\e_{\b\a}F(0,0).}\\
&\im\{\vr^\a Q_\a,\psi_b(0,0)\}=i\e_{\b\a}\vr^\a F(0,0),\\
&\im \boxed{\de_{\vr}\psi_\b (0,0)=i\vr_\b F(0,0).}
\end{aligned}
\ee
  
\be{108}
\begin{aligned}
&\bullet \{Q_{\a},\bQ_{\da},\psi_\b(0,0)\}:\\
&\implies [\{Q_{\a},\bQ_{\da}\},\psi_\b]+[\{\psi_\b,\bQ_{\da}\},Q_{\a}]+[\{\psi_\b,Q_{\a}\},\bQ_{\da}]=0,\\
&\implies{[2(\s^0)_{\a\da}H,\psi_\b(0,0)]}-2(\s^0)_{\b\da}\p_t\psi_\a(0,0)=i\e_{\b\a}[\bQ_{\da},F(0,0)],\\
&\im 2i[\bQ_{\da},F(0,0)]=2(\s^0)_{\a\da}\e^{\a\b}\p_t\psi_\b(0,0)-2(\s^0)_{\b\da}\e^{\a\b}\p_t\psi_\a(0,0),~~\text{(using, }\e_{\a\b}\e^{\b\a}=2)\\
&\im\boxed{[\bQ_{\da},F(0,0)]=-2i(\p_t\psi\;\s^0)_{\da}},\\
&\im[\bQ_{\da}\bvr^{\da},F(0,0)]=-2i(\p_t\psi\;\s^0)_{\da}\bvr^{\da},\\
&\im\boxed{\bar{\de}_{\bvr} F(0,0)=-2i(\p_t\psi\;\s^0\;\bvr)}.
\end{aligned}
\ee

\be{leftF2}
 \hspace{-5cm}
\begin{aligned}
&\bullet \{Q_{\a},\bQ_{\da},F(0,0)\}:\\
&\implies [\{Q_{\a},\bQ_{\da}\},F]+\{[F,Q_{\a}],\bQ_{\da}\}+\{[F,\bQ_{\da}],Q_{\a}\}=0,\\
&\implies{[2(\s^0)_{\a\da}H,F]}+\{2i(\p_t\psi\s^0)_{\da},Q_\a\}=\{\bQ_{\da},[Q_\a,F]\},\\
&\im \{\bQ_{\da},[Q_\a,F]\}=2\((\s^0)_{\a\da}-\e^{\rho\b}\e_{\b\a}(\s^0)_{\rho\da}\)\p_t F,\\
&\hspace{3.4cm}=2\((\s^0)_{\a\da}-\de^\rho_{\a}(\s^0)_{\rho\da}\)\p_t F,\\
&\im  \{\bQ_{\da},[Q_\a,F]\}=0,\\
&\im \boxed{[Q_\a,F(0,0)]=0,~~\de_{\vr}F(0,0)=0.}
\end{aligned}
\ee
The equations in boxes give infinitesimal transformation of the fields under the fermionic generators $Q, \bQ$. The above transformation relations of the fields $\{\phi, \psi,F\}$ are consistent with Jacobi's involving all other finite CSA generators.

\medskip

\paragraph{Checking for consistency:}
We check the consistency of the above infinitesimal transformations. We are considering a generic field $\Phi$ here. Then
\be{infinl1}
 \hspace{-7cm}
\begin{aligned}
&\bullet\{Q_{\a},\bQ_{\db}\}=2(\s^0)_{\a\db}H,\\
&\im\vr_1^{\a}(Q_{\a}\bQ_{\db}+\bQ_{\db}Q_{\a}){\bvr}^{\db}_2=2\vr_1^{\a}(\s^0)_{\a\db}{\bvr}^{\db}_2H,\\
&\im\boxed{ [\de_{\vr_1},\bar{\de}_{\bvr_2}]\Phi=2(\vr_1\:(\s^0)\;{\bvr}_2)\p_t\Phi.}
\end{aligned}
\ee
We have rearranged the fermionic objects and changed the signs appropriately in reaching the last step.
Similarly,
\be{infinl2}
 \hspace{-8cm}
\begin{aligned}
&\bullet\{Q_{\a},Q_{\b}\}=0,~~\im \boxed{[\de_{\vr_1},{\de}_{\vr_2}]\Phi=0.}\\
&\bullet\{\bQ_{\da},\bQ_{\db}\}=0,~~\im \boxed{[\bar{\de}_{{\bvr}_1},\bar{\de}_{{\bvr}_2}]\Phi=0.}
\end{aligned}
\ee
The infinitesimal transformations mentioned in boxes in equations \eqref{leftphi1},\eqref{leftphi2},\eqref{leftpsi1},\eqref{leftpsi2}, \eqref{leftF1}, \eqref{leftF2} should respect \eqref{infinl1} and \eqref{infinl2}. We check this explicitly below for each field. 
\be{109}
 \hspace{-1cm}
\begin{aligned}
\bullet [\de_{\vr_1},\bde_{{\bvr}_2}]\phi&=0-\bde_{{\bvr}_2}\de_{\vr_1}\phi,=-\de_{\bvr_2}(i\vr_1\psi)=-i\vr^{\b}_1(2i\s^0\bvr_2)_{\b}\p_t\phi,=2(\vr_1\s^0\bvr_2)\p_t\phi^\star.
\end{aligned}
\ee
\be{110}
 \hspace{-0.5cm}
\begin{aligned}
\bullet [\de_{\vr_1},\bde_{{\bvr}_2}]F&=\de_{\vr_1}\bde_{{\bvr}_2}F-0=\de_{{\vr}_1}(-2i\p_t\psi\s^0\bvr_2)=-2i\e^{\a\b}\de_{{\vr}_1}(\p_t\psi_\b)(\s^0\bvr_2)_\a=2\vr^\a_1(\s^0\bvr_2)_\a\p_tF\\
&=2(\vr_1\s^0\bvr_2)\p_t F.
\end{aligned}
\ee

\be{psiconst}
\hspace{-3cm}
\begin{aligned}
\bullet [\de_{\vr_1},\bde_{{\bvr}_2}]\psi_{\b}&=\de_{\vr_1}\bde_{{\bvr}_2}\psi_{\b}-\bde_{{\bvr}_2}\de_{\vr_1}\psi_{\b}=-2(\s^0\bvr_2)_{\b}(\vr_1\p_t\psi)-2{\vr}_{1_{\b}}(\p_t\psi\s^0\bvr_2)
\end{aligned}
\ee
To evaluate the terms in \eqref{psiconst}, we use the Fierz identities. For general spinors,

\be{fierzleft1}
 \hspace{-5.5cm}
\begin{aligned}
\bullet~ \bar{\lambda}^{\da}(\phi\cdot\chi)&=-\phi^\rho\bar{\lambda}^{\da}\chi_\rho=-\phi^\b(\de^{\rho}_{\b}\de^{\da}_{\dr})\bar{\lambda}^{\rho}\chi_\rho\\
&=-\phi^\b\(\frac{1}{2}\)\left[(\bs^0)^{\da\rho}(\s^0)_{\b\dr}-(\bs^i)^{\da\rho}(\s^i)_{\b\dr}\right]\bar{\lambda}^{\rho}\chi_\rho\\
&=-\(\frac{1}{2}\)({\phi}\s^0 \bar{\lambda})( \bs^0\chi)^{\da}+\(\frac{1}{2}\)({\phi}\s^i \bar{\lambda})( \bs^i\chi)^{\da}
\end{aligned}
\ee
We have used the following identity to evaluate \eqref{fierzleft1}.
\be{111}
\de^{\dr}_{\da}\de^{\rho}_{\a}=\(\frac{1}{2}\)\left[(\bs^0)^{\dr\rho}(\s^0)_{\a\da}-(\bs^i)^{\dr\rho}(\s^i)_{\a\da}\right]
\ee
Similarly,
\bea{fierzleft2}
 \hspace{-2cm}\non\bullet~ {\phi}_{\a}(\bar{\lambda}\cdot\bar{\chi})&&=-\bar{\lambda}_{\dr}(\de^{\dr}_{\db}\de^{\rho}_{\a})\phi_\rho\bar{\chi}^{\db}=-\bar{\lambda}_{\da}\(\frac{1}{2}\)\left[(\bs^0)^{\dr\rho}(\s^0)_{\a\db}-(\bs^i)^{\dr\rho}(\s^i)_{\a\db}\right]{\phi}^{\b}\chi^{\db}\\
 \hspace{-10cm}&&=-\(\frac{1}{2}\)(\bar{\lambda}\bs^0{\phi} )( \s^0\bar{\chi})_{\a}+\(\frac{1}{2}\)(\bar{\lambda}\bs^i{\phi} )( \s^i\bar{\chi})_{\a}
\eea
Using \eqref{fierzleft1} in the first term of \eqref{psiconst},
\bea{d.19}
\non-2(\s^0\bvr_2)_{\b}(\vr_1\p_t\psi)&&=(\s^0)_{\b\db}[(\vr_1 \s^0 \bvr_2)(\bs^0\p_t\psi )^{\db}-(\vr_1 \s^i \bvr_2)(\bs^i\p_t\psi )^{\db}]\\
&&=(\vr_1 \s^0 \bvr_2)( \s^0 \bs^0\p_t\psi)_\b-(\vr_1 \s^i \bvr_2)( \s^0 \bs^i\p_t\psi)_\b\,.
\eea
Also, using \eqref{fierzleft2} in the second term of \eqref{psiconst},
\bea{d.20}
\non-2{\vr}_{1_{\b}}(\p_t\psi\s^0\bvr_2)&&=2{\vr}_{1_{\b}}(\bvr_2\cdot\bs^0\p_t\psi)\\
\non&&=-(\bvr_2\bs^0\vr_1)(\s^0\bs^0\p_t\psi)_{\b}+(\bvr_2\bs^i\vr_1)(\s^i\bs^0\p_t\psi)_{\b}\\
&&=(\vr_1\s^0\bvr_2)(\p_t\psi)_{\b}-(\vr_1\s^i\bvr_2)(\s^i\bs^0\p_t\psi)_{\b}\,.
\eea
Thus, adding \eqref{d.19} and \eqref{d.20},
\be{112}
\begin{aligned}
{[\de_{\vr_1},\bde_{{\bvr}_2}]\psi_{\b}}&=2(\vr_1\s^0\bvr_2)(\p_t\psi)_{\b}-(\vr_1\s^i\bvr_2)[(\s^0)_{\b\dr}(\bs^i)^{\dr\ga}+(\s^i)_{\b\dr}(\bs^0)^{\dr\ga}](\p_t\psi)_{\ga}\\
&=2(\vr_1\s^0\bvr_2)(\p_t\psi)_{\b}~.
\end{aligned}
\ee
It satisfies \eqref{infinl1}. Hence, the transformation relations of the fields $\{\phi, \psi, F\}$ under $Q,\bQ$ are consistent.

\paragraph{Transformation under $\boldsymbol{R}$ symmetry generator:} Now, we discuss the transforamtion under $R$ symmetry generators in \eqref{73.5}.
The states are labelled under $R$ charges.
Let,
\be{113}
{[R,\phi]=r\phi}\,.
\ee
Next, we show the Jacobi identities which help us find the transformation relations. 
\be{114}
 \hspace{-7cm}
\begin{aligned}
&\bullet \{Q_\a,R,\phi\}:\\
&\im [Q_\a,[R,\phi]]+[R,[\phi,Q_\a]]+[\phi,[Q_\a,R]]=0,\\
&\im[Q_\a,[R,\phi]]=[R,i\psi_\a]+[\frac{3i}{4}Q_\a,\phi],\\
&\im\boxed{[R,\psi_\a]=(r-\frac{3i}{4})\psi_\a.}
\end{aligned}
\ee
\be{115}
 \hspace{-6cm}
\begin{aligned}
&\bullet \{Q_\a,\psi_\b,R\}:\\
&\im[\{Q_\a,\psi_\b\},R]+\{[R,\psi_\b],Q_\a\}+\{[R,Q_\a],\psi_\b\}=0,\\
&\im \e_{\b\a}[R,F]=(r-\frac{3i}{4})\e_{\b\a}F-\frac{3i}{4}\e_{\b\a}F,\\
&\im \boxed{[R,F]=(r-\frac{3i}{2})F.}
\end{aligned}
\ee
The boxed equations in \eqref{113}--\eqref{115} give the required relations.
\paragraph{$\boldsymbol{S}$ transformations on the fields:} Here, we will discuss details of the $S$ transformations on the fields mentioned in \eqref{leftS1}.
For a CSA primary $\Phi$,
\be{116}
[S_\a,\Phi(0,0)]=[{\bS}_{\da},\Phi(0,0)]=0\,.
\ee
\be{117}
[\mathcal{O},\Phi(x_i,t)]=U[U^{-1}\mathcal{O} U,\Phi(0,0)]U^{-1},~~U=e^{-tH-x_iP_i}
\ee
We find,
\bes{}
\label{d28}
\begin{eqnarray}
\hspace{-6cm}&&U^{-1}Q_{\a}U=Q_{\a}, ~~ (\text{as,~}[H,Q_{\a}]=[P_i,Q_{\a}]=0),\\
\hspace{-6cm}&&U^{-1}\bQ_{\da}U=\bQ_{\da}, ~~ (\text{as,~}[H,\bQ_{\da}]=[P_i,\bQ_{\da}]=0),\\
\hspace{-6cm}&&U^{-1}S_{\a}U=S_{\a}-i(x_i \sigma_i)_{\a\da}\:\e^{\da\db}{\bQ}_{\db},\\
\hspace{-6cm}&&U^{-1}{\bS}_{\da}U={\bS}_{\da}-i \e_{\da\db}(x_i \bar{\sigma}_i)^{\db\a}\:Q_{\a}\,.
\end{eqnarray}
\ees
We put \eqref{d28} in \eqref{117} to write
\bes\label{leftS}
\begin{eqnarray}
\hspace{-6cm}&&[S_\a,\phi(t,x_i)]=0,\\
\hspace{-6cm}&&[\bS_{\da},\phi(t,x_i)]=\e_{\da\db}(x_i \bar{\sigma}_i)^{\db\a}\:\psi_{\a}(t,x_i),\\
\hspace{-6cm}&&\{S_\a,\psi_{\b}(t,x_i)\}=-2(x_i \sigma_i)_{\a\da}\:\e^{\da\db}(\bar{\s}^0)_{\db\b}\p_t\phi(t,x_i),\\
\hspace{-6cm}&&\{{\bS}_{\da},\psi_{\b}(t,x_i)\}=-\e_{\da\db}(x_i \bar{\sigma}_i)^{\db\a}\e_{\a\b}F(t,x_i),\\
\hspace{-6cm}&&[{\bS}_{\db},F(t,x_i)]=0,\\
\hspace{-6cm}&&[S_\b,F(t,x_i)]=-2(x_i \sigma_i)_{\b\db}\:\e^{\db\dg}(\bar{\s}^0)_{\dg\rho}\p_t\psi^{\rho}(t,x_i)\,.
\end{eqnarray}
\ees

\subsection{Right chiral sector}\label{representation2}
The transformation relations of the fields $\{\phi^{\star},\chi,F^{\star}\}$ in the right chiral sector under finite fermionic generators $Q,\bQ,S,\bS$  are given by \eqref{540}--\eqref{rightS}. We look into the details of the calculations here.
First, we consider the Jacobi Identities for the right chiral sector mentioned in \eqref{540a}--\eqref{540c}.
\be{rightphi1}
\begin{aligned}
&\bullet\{B_i,\phi^\star(0,0),\bQ_{\da}\}:\\
&\implies [[B_i,\phi^\star],\bQ_{\da}]+[[\phi^\star,\bQ_{\da}],B_i]+[[\bQ_{\da},B_i],\phi^\star]=0,\\
&\implies {[[\bQ_{\da},\phi^\star],B_i]}=0,\\
&\implies\boxed{ [\bQ_{\da},\phi^\star(0,0)]=i\chi_{\da}(0,0).}\\
&\text{Also,~} [{\bvr}^{\da} \bQ_{\da},\phi^\star(0,0)]=i{\bvr^{{\da}}}\chi_{\da}(0,0),\\
&\implies\boxed{\bar{\delta}_{\bvr}\phi^\star(0,0)=i{\bvr}\chi(0,0).}~~~~~~~\text{(where,~}{{\bvr}\chi}={\bvr}_{{\da}}\chi^{{\da}}=-{\bvr}^{{\da}}\chi_{{\da}},~\bar{\delta}_{\bvr}={\bvr}_{\da} \bQ^{\da}=-{\bvr}^{\da} \bQ_{\da},).
\end{aligned}
\ee

\be{rightphi2}
\hspace{-7cm}
\begin{aligned}
&\bullet\{B_i,\phi^\star(0,0),Q_{\a}\}:\\
&\implies {[[Q_{\a},\phi^\star],B_i]}=0,\implies \boxed{[Q_{\a},\phi^\star(0,0)]=0.}\\
&\text{Also,~}  [{\vr}^{\a}Q_{\a},\phi^\star(0,0)]=0,\implies \boxed{\de_{\vr} \phi^\star(0,0)=0.}
\end{aligned}
\ee

\be{118}
\hspace{-6cm}
\begin{aligned}
&\bullet\{J_{ij},\phi^\star(0,0),\bQ_{\da}\}:\\
&\implies{[[J_{ij},\phi^\star],\bQ_{\da}]}+[[\phi^\star,\bQ_{\da}],J_{ij}]+[[\bQ_{\da},J_{ij}],\phi^\star]=0,\\
&\implies -\frac{i}{2}\e_{\da\db}(\bar{\s}^{ij})^{\db}_{\:\:\dot{\rho}}\e^{\dot{\rho}\dg}[\bQ_{\dg},\phi^\star]+[[\phi^\star,\bQ_{\da}],J_{ij}]=0,\\
&\implies i[J_{ij},\chi_{\da}(0,0)]=\frac{i}{2}\e_{\da\db}(\bar{\s}^{ij})^{\db}_{\:\:\dot{\rho}}\e^{\dot{\rho}\dg}[\bQ_{\dg},\phi^\star(0,0)]\\
&\implies\boxed{[J_{ij},\chi_{\da}(0,0)]=\frac{i}{2}\e_{\da\db}(\bar{\s}^{ij})^{\db}_{\:\:\dot{\rho}}\e^{\dot{\rho}\dg}\chi_{\dg}(0,0).}
\end{aligned}
\ee

\be{leftpsi11}
\hspace{-1cm}
\begin{aligned}
&\bullet \{Q_{\a},{\bQ}_{\da},\phi^\star(0,0)\}:\\
&\implies [\{Q_{\a},\bQ_{\da}\},\phi^\star]+\{[\phi^\star,Q_{\a}],\bQ_{\da}\}+\{[\phi^\star,\bQ_{\da}],Q_{\a}\}=0,\\
&\implies{[2(\s^0)_{\a\da}H,\phi^\star]}+i\{\chi_{\da},Q_{\a}\}=0,\\
&\implies\boxed{\{Q_{\a},\chi_{\da}\}=-2i(\s^0)_{\a\da}\p_t\phi^\star(0,0).}\\
&\text{Also,~} [ \vr^{\a}Q_{\a},\chi_{\db}(0,0)]=-2i\vr^{\a}(\s^0)_{\a\db}\p_t\phi^\star(0,0),~~~~~~\text{(using,~}\vr^{\a}Q_{\a}=-\vr_{\a}Q^{\a}={\de}_{\vr})\\
&\implies\boxed{\delta_{\vr}\chi_{\db}(0,0)=-2i(\vr\s^0 )_{\db}\p_t \phi^\star(0,0).}~~~~~~~~~~~~~~~~~~\text{(where,~}(\vr\s^0 )_{\db}=\vr^{\a}(\s^0)_{\a\db})
\end{aligned}
\ee

\be{d.33}
\hspace{-1cm}
\begin{aligned}
&\bullet  \{J_{ij},\bQ_{\da},\chi_{\db}(0,0)\}: \\
&\im [\{\bQ_{\da},\chi_{\db}\},J_{ij}]+\{[J_{ij},\chi_{\db}],\bQ_{{\da}}\}+\{[J_{ij},\bQ_{{\da}}],\chi_{\db}\}=0,\\
&\im [J_{ij},\{\bQ_{\da},\chi_{\db}(0,0)\}]=\{[J_{ij},\chi_{\db}],\bQ_{\da}\}+\{[J_{ij},\bQ_{\da}],\chi_{\db}\},\\
&~~~~~~~~~~~~~~~~~~= \frac{i}{2}\left(\e_{\db\dr}(\bs_{ij})^{\dr}_{\;\;\dot{\s}}\e^{\dot{\s}\dg}\{\chi_{\dg}(0,0),\bQ_{\da}\}+\e_{\da\dr}(\bs_{ij})^{\dr}_{\;\;\dot{\s}}\e^{\dot{\s}\dg}\{\bQ_{\dg},\chi_{\db}(0,0)\} \right),\\
\end{aligned}
\ee
The LHS of \eqref{d.33} should be zero, as $\{\bQ_{\da},\chi_{\db}\}$ gives a scalar. We can check that the RHS also gives zero if,
\be{119}
\{\bQ_{\da},\chi_{\db}(0,0)\}=-i\e_{\db{\da}}F^\star(0,0).
\ee

Hence,
\be{rightF1}
\begin{aligned}
&\boxed{\{\bQ_{\da},\chi_{\db}(0,0)\}=-i\e_{\db{\da}}F^\star(0,0).}\\
&\im\{\bvr^{\da} \bQ_{\da},\chi_{\db}(0,0)\}=-i\e_{\db{\da}}\vr^{\da} F^\star(0,0),\\
&\im \boxed{\bde_{\bvr}\chi_{\db} (0,0)=i\bvr_{\db} F^\star(0,0).}
\end{aligned}
\ee

\be{120}\hspace{-0.5cm}
\begin{aligned}
&\bullet \{Q_{{\a}},\bQ_{\da},\chi_{\db}(0,0)\}:\\
&\implies [\{Q_{{\a}},\bQ_{\da}\},\chi_{\db}]+[\{\chi_{\db},Q_{\a}\},\bQ_{{\da}}]+[\{\bQ_{{\da}},\chi_{\db}\},Q_{\a}]=0,\\
&\implies{[2(\s^0)_{{\a}\da}H,\chi_{\db}(0,0)]}-2(\s^0)_{\a{\db}}\p_t\chi_{\da}(0,0)=-i\e_{{\db}{\da}}[Q_{\a},F^\star(0,0)],\\
&\im -2i[Q_{\a},F^\star(0,0)]=2(\s^0)_{{\a}\da}\e^{{\da}{\db}}\p_t\chi_{\db}(0,0)-2(\s^0)_{\a{\db}}\e^{{\da}{\db}}\p_t\chi_{\da}(0,0),~~\text{(using, }\e_{{\da}{\db}}\e^{{\db}{\da}}=2)\\
&\im\boxed{[Q_{\a},F^\star(0,0)]=2i(\s^0)_{\a\da}\e^{\da\dr}\p_t\chi_{\dr},}\\
&\im[\vr^{\a}Q_{\a},F^\star(0,0)]=2i\vr^{\a}(\s^0)_{\a\da}\e^{\da\dr}\p_t\chi_{\dr},\\
&\im\boxed{{\de}_{\vr} F^\star(0,0)=2i(\vr\:\s^0\p_t\chi).}
\end{aligned}
\ee

\be{rightF^2}
\hspace{-5cm}
\begin{aligned}
&\bullet \{Q_{\a},\bQ_{\da},F^{\star}(0,0)\}:\\
&\implies [\{Q_{\a},\bQ_{\da}\},F^{\star}]+\{[F^{\star},Q_{\a}],\bQ_{\da}\}+\{[F^{\star},\bQ_{\da}],Q_{\a}\}=0,\\
&\implies{[2(\s^0)_{\a\da}H,F^{\star}]}-2i(\s^0)_{\a\db}\e^{\db\dr}\{\p_t\chi_{\dr},Q_\a\}=\{[\bQ_{\da},F^{\star}],Q_\a\},\\
&\im\{[\bQ_{\da},F^{\star}],Q_\a\}=2\((\s^0)_{\a\da}-(\s^0)_{\a\db}\e^{\db\dr}\e_{\dr\da}\)\p_t F^{\star},\\
&\hspace{3.4cm}=2\((\s^0)_{\a\da}-(\s^0)_{\a\db}\de^{\db}_{\da}\)\p_t F^{\star},\\
&\im  \{[\bQ_{\da},F^{\star}],Q_\a\}=0,\im \boxed{[\bQ_{\da},F^{\star}]=0,~~\bde_{\bvr}F^{\star}(0,0)=0.}
\end{aligned}
\ee

\paragraph{Checking for consistency:} Now, we check the consistency of the transformation relations found in the boxed equations of \eqref{rightphi1}--\eqref{122}. They should obey the consistency relations in \eqref{d.20}--\eqref{112}.  We are showing here only the consistency relation in \eqref{d.20}, while the other ones are also satisfied.
\be{121}
\hspace{-0.5cm}
\begin{aligned}
\bullet [\de_{\vr_1},\bde_{{\bvr}_2}]\phi^\star&=\de_{\vr_1}\bde_{{\bvr}_2}\phi^\star-0,=\de_{\vr_1}(i\bvr_2\chi),=-2i^2(\vr_1\s^0)_{\db}\bvr^{\db}_2\p_t\phi^\star,=2(\vr_1\s^0\bvr_2)\p_t\phi^\star.
\end{aligned}
\ee
\be{122}
\begin{aligned}
\bullet [\de_{\vr_1},\bde_{{\bvr}_2}]F^\star&=-\bde_{{\bvr}_2}\de_{\vr_1}F^\star,=-\bde_{{\bvr}_2}(2i(\vr_1 \s^0)_{\dr}\e^{\dr\db}\p_t\chi_{\db}),=2(\vr_1\s^0)_{\dr}\e^{\dr\db}\bvr_{2_{\db}}\p_t F^\star,=2(\vr_1\s^0\bvr_2)\p_t F^\star.
\end{aligned}
\ee
\bea{chiconst}
\hspace{-3cm}\bullet [\de_{\vr_1},\bde_{{\bvr}_2}]\chi_{\db}&&=\de_{\vr_1}\bde_{{\bvr}_2}\chi_{\db}-\bde_{{\bvr}_2}\de_{\vr_1}\chi_{\db}=-2{\bvr}_{2_{\db}}(\vr_1\s^0\p_t\chi)-2(\vr_1\s^0)_{\db}(\p_t\chi\bvr_2)
\eea
To evaluate the terms in \eqref{chiconst}, we use the Fierz identities. For general spinors,
\be{fierzright1}
\hspace{-5cm}
\begin{aligned}
\bullet~ \bar{\phi}_{\da}(\lambda\cdot\chi)&=-\lambda^{\a}\bar{\phi}_{\da}\chi_{\a}=-\lambda^{\a}\bar{\phi}_{\dr}(\de^{\dr}_{\da}\de^{\rho}_{\a})\chi_{\rho}\\
&=-\lambda^{\a}\bar{\phi}_{\dr}\(\frac{1}{2}\)\left[(\bs^0)^{\dr\rho}(\s^0)_{\a\da}-(\bs^i)^{\dr\rho}(\s^i)_{\a\da}\right]\chi_{\rho}\\
&=-\(\frac{1}{2}\)(\bar{\phi}\bs^0 \chi)(\lambda \s^0)_{\da}+\(\frac{1}{2}\)(\bar{\phi}\bs^i \chi)(\lambda \s^i)_{\da}
\end{aligned}
\ee
We have used the following identity to evaluate \eqref{fierzright1}.
\be{123}
\non\de^{\dr}_{\da}\de^{\rho}_{\a}=\(\frac{1}{2}\)\left[(\bs^0)^{\dr\rho}(\s^0)_{\a\da}-(\bs^i)^{\dr\rho}(\s^i)_{\a\da}\right]
\ee
Similarly,
\be{fierzright2}
\hspace{-5cm}
\begin{aligned}
\bullet~ {\phi}^{\a}(\bar{\lambda}\cdot\bar{\chi})&=-\bar{\lambda}_{\da}{\phi}^{\rho}(\de^{\a}_{\rho}\de^{\da}_{\db})\bar{\chi}^{\db}\\
&=-\bar{\lambda}_{\da}{\phi}^{\rho}\(\frac{1}{2}\)\left[(\bs^0)^{\da\a}(\s^0)_{\rho\db}-(\bs^i)^{\da\a}(\s^i)_{\rho\db}\right]\chi^{\db}\\
&=-\(\frac{1}{2}\)({\phi}\s^0 \bar{\chi})(\bar{\lambda} \bs^0)^{\a}+\(\frac{1}{2}\)({\phi}\s^i \bar{\chi})(\bar{\lambda} \bs^i)^{\a}
\end{aligned}
\ee
Using \eqref{fierzright2} in the second term of \eqref{chiconst},
\be{124}
\begin{aligned}
-2(\vr_1\s^0)_{\db}(\p_t\chi\bvr_2)&=[(\vr_1 \s^0 \bvr_2)(\p_t\chi \bs^0)^\b-(\vr_1 \s^i \bvr_2)(\p_t\chi \bs^i)^\b](\s^0)_{\b\db}\\
&=(\vr_1 \s^0 \bvr_2)(\p_t\chi \bs^0 \s^0)-(\vr_1 \s^i \bvr_2)(\p_t\chi \bs^i \s^0).
\end{aligned}
\ee
Also, using \eqref{fierzright1} in the first term of \eqref{chiconst},
\be{125}
\begin{aligned}
-2{\bvr}_{2_{\db}}(\vr_1\s^0\p_t\chi)&=2{\bvr}_{2_{\db}}(\p_t\chi\bs^0\bvr_1)\\
&=-(\bvr_2\bs^0\vr_1)(\p_t\chi\bs^0\s^0)_{\db}+(\bvr_2\bs^i\vr_1)(\p_t\chi\bs^0\s^i)_{\db}\\
&=(\vr_1\s^0\bvr_2)(\p_t\chi)_{\db}-(\vr_1\s^i\bvr_2)(\p_t\chi\bs^0\s^i)_{\db}.
\end{aligned}
\ee
Thus,
\be{126}
\begin{aligned}
{[\de_{\vr_1},\bde_{{\bvr}_2}]\chi_{\db}}&=2(\vr_1\s^0\bvr_2)(\p_t\chi)_{\db}-(\vr_1\s^i\bvr_2)(\p_t\chi)_{\dr}[(\bs^i)^{\dr\s}(\s^0)_{\s\db}+(\bs^0)^{\dr\s}(\s^i)_{\s\db}]\\
&=2(\vr_1\s^0\bvr_2)(\p_t\chi)_{\db}.
\end{aligned}
\ee
Hence, the transformation relations in \eqref{rightphi1}--\eqref{122} are consistent.

\paragraph{Transformation under $\boldsymbol{R}$ symmetry generator:}
Let us focus on equations \eqref{544}, which give the transformations under $R$ symmetry generators. The states are labelled under $R$ charges.
Let
\be{127}
\boxed{{[R,\phi^{\star}]=r^\star\phi^{\star}}.}
\ee
\be{128}
\hspace{-6.5cm}
\begin{aligned}
&\bullet \{\bQ_{\da},R,\phi^{\star}\}:\\
&\im [\bQ_{\da},[R,\phi^{\star}]]+[R,[\phi^{\star},\bQ_{\da}]]+[\phi^{\star},[\bQ_{\da},R]]=0,\\
&\im[\bQ_{\da},r^\star \phi^\star]+[R,-i\chi_{\da}]+[\phi^{\star},-\frac{3i}{4}\bQ_{\da}]=0,\\
&\im\boxed{[R,\chi_{\da}]=(r^\star+\frac{3i}{4})\chi_{\da}.}
\end{aligned}
\ee
\be{129}
\hspace{-6cm}
\begin{aligned}
&\bullet \{\bQ_{\da},\chi_{\db},R\}:\\
&\im[\{\bQ_{\da},\chi_{\db}\},R]+\{[R,\chi_{\db}],\bQ_{\da}\}+\{[R,\bQ_{\da}],\chi_{\db}\}=0,\\
&\im \e_{{\db}{\da}}[F^{\star},R]+(r^\star+\frac{3i}{4})\e_{{\db}\da}F^{\star}+\frac{3i}{4}\e_{{\db}\da}F^{\star}=0,\\
&\im \boxed{[R,F^{\star}]=(r+\frac{3i}{2})F^{\star}.}
\end{aligned}
\ee
Finally, for the $S$ transformations we repeat the calculation similar to the left chiral sector to obtain,
\bes\label{rightS1}
\begin{eqnarray}
\hspace{-6cm}&&[S_\a,\phi^\star(t,x_i)]=(x_i {\sigma}_i)_{\a\da}\e^{\da\db}\:\chi_{\db}(t,x_i),\\
\hspace{-6cm}&&[\bS_{\da},\phi^\star(t,x_i)]=0,\\
\hspace{-6cm}&&\{S_\a,\chi_{\db}(t,x_i)\}=-(x_i {\sigma}_i)_{\a\db}F^\star(t,x_i),\\
\hspace{-6cm}&&\{{\bS}_{\da},\chi_{\db}(t,x_i)\}=-2\e_{\da\db}(x_i \bar{\sigma}_i)^{\db\a}\:({\s}^0)_{\a\dr}\p_t\phi^\star(t,x_i),\\
\hspace{-6cm}&&[{\bS}_{\da},F^\star(t,x_i)]=2\e_{\da\db}(x_i \bar{\sigma}_i)^{\db\a}\:({\s}^0)_{\a\dr}\p_t\chi^{\dr}(t,x_i),\\
\hspace{-6cm}&&[S_\a,F^\star(t,x_i)]=0\,.
\end{eqnarray}
\ees

\bibliographystyle{JHEP}
\bibliography{V3_carroll_scft}

\end{document}